\documentclass[a4paper,11pt]{article}

\usepackage{amsbsy}
\usepackage{amsfonts}
\usepackage{amsmath}
\usepackage{amssymb}
\usepackage[titletoc,toc,title]{appendix}
\usepackage{authblk}
\usepackage{braket}
\usepackage{cancel}
\usepackage{caption}
\usepackage{cite}
\usepackage{color}
\usepackage{comment}
\usepackage{enumitem}
\usepackage{epsfig}
\usepackage{etoolbox}
\usepackage{float}
\usepackage[margin=2.5cm]{geometry}
\usepackage{graphicx}
\usepackage{hyperref}
\usepackage{cleveref} 
\usepackage[none]{hyphenat}
\usepackage{mathtools}
\usepackage{ragged2e}
\usepackage{setspace}
\usepackage{soul}
\usepackage{subcaption}
\usepackage{tikz}
\usepackage{titlesec}
\usepackage[english]{babel}

\makeatletter
\patchcmd{\@maketitle}{center}{flushleft}{}{}
\patchcmd{\@maketitle}{center}{flushleft}{}{}
\patchcmd{\@maketitle}{\LARGE}{\LARGE\sffamily\bfseries}{}{}
\makeatother

\makeatletter
\def\maketitle{{%

\AB@maketitle}}

\renewcommand\AB@affilsepx{\protect\\\vskip1em\protect\Affilfont}

\makeatother

\renewcommand\Affilfont{\normalsize\normalfont\itshape\small}

{\makeatletter\g@addto@macro\bfseries{\boldmath}\makeatother}

\makeatletter
\g@addto@macro\ps@plain{%
\def\@oddfoot{\reset@font\hfill-- \thepage\ --\hfill}%
\let\@evenfoot\@oddfoot
}
\makeatother
\titleformat*{\section}{\normalfont\large\bfseries}
\titleformat*{\subsection}{\normalfont\normalsize\bfseries}
\titleformat*{\subsubsection}{\normalfont\normalsize\bfseries}
\titleformat*{\paragraph}{\normalfont\normalsize\bfseries}
\titleformat*{\subparagraph}{\normalfont\normalsize\bfseries}

\usetikzlibrary{shapes}

\hypersetup{%
citecolor=red,
colorlinks=true,
filecolor=red,
linkcolor=blue,
linktocpage=true,
urlcolor=cyan
}

\begin{document}

\title{Entanglement at the Soft-Hair Horizon
\vskip23pt plus0.06fil minus13pt
\vskip -5pt
\hrule height 1.5pt
\vskip23pt plus0.06fil minus13pt
\vskip -25pt
}

\author[1]{Sayid Mondal\thanks{\noindent E-mail:~ {\tt sayid.mondal@gmail.com}}}
\author[1,2]{Wen-Yu Wen\thanks{\noindent E-mail:~ {\tt  wenw@cycu.edu.tw}}}

\affil[1]{%
\parbox[t]{\linewidth}{%
Center for High Energy Physics and Department of Physics\\

Chung-Yuan Christian University, Taoyuan, Taiwan
}
}
\affil[2]{%
\parbox[t]{\linewidth}{%
Leung Center for Cosmology and Particle Astrophysics\\

National Taiwan University, Taipei, Taiwan
}
}

\date{}

\begin{NoHyper}
\maketitle
\end{NoHyper}

\thispagestyle{empty}

\begin{abstract}\noindent
\justify

We study entanglement between two modes of a free bosonic and fermionic field as seen by two relatively accelerating observers at the soft-hair horizon of a four-dimensional supertranslated Schwarzschild black hole.  First, we associate the shock wave to the near horizon geometry by explicit construction of map between their metric as well as relation between wave form factor and soft hair function.  We then compute mutual information and entanglement negativity as a measure of entanglement between these two observers and show that their angular dependence at the soft-hair horizon. We also show that for vanishing soft hair function our results are equal to that of an ordinary Schwarzschild black hole case.

\end{abstract}

\clearpage
\hrule
\tableofcontents
\bigskip\medskip
\hrule
\bigskip\bigskip

%

\setlength\parindent{1.2\parindent}
\setstretch{1.1}

\section{Introduction}
\label{sec_intro}
Over the last few decades entanglement has evolved as a central attention in the
study of diverse physical phenomena ranging from quantum many-body systems to the process of black hole formation and various other issues in quantum gravity. It is well known that for a bipartite $(A\cup B)$ system in a pure state $\rho$, the entanglement entropy  characterizes the amount of entanglement which is given by the von-Neumann entropy of the reduced density matrix $\rho_A=\mathrm{Tr}_{B}\rho$ of the subsystem $A$ as
\begin{equation}
S_A=- \mathrm{Tr}(\rho_A \log \rho_A).
\end{equation}  
Another important quantity that is constructed from the algebraic sum of entanglement entropies known as {\it mutual information}:
\begin{equation}\label{muinfo}
I(A:B)=S_A+ S_B-S_{AB},
\end{equation}
that is symmetric in subsystems $A$ and $B$. It is important to note that $I(A : B)$ has not all the properties to be an entanglement
measure, but it provides an upper bound on the total amount of correlations
between the subsystems $A$ and $B$.  In quantum field theory (QFT) entanglement entropy is obtained by a {\it replica trick} developed by the authors in \cite{Calabrese:2004eu,Calabrese:2005zw,Calabrese:2009qy}. For holographic characterization of the entanglement entropy in dual conformal field theories (CFTs) see \cite{Ryu:2006bv,Ryu:2006ef,Takayanagi:2012kg,Nishioka:2009un,Nishioka:2018khk,e12112244,Blanco:2013joa,Fischler2013,Fischler:2012uv,Chaturvedi:2016kbk,Fursaev:2006ih,Headrick:2010zt,Faulkner:2013yia,Casini:2011kv,Lewkowycz:2013nqa,Hubeny:2007xt,Dong:2016hjy} (and references therein). 

It is well known that entanglement entropy fails to characterize mixed state entanglement as it involves correlations irrelevant to the entanglement of the specific mixed state. To resolve this intricate issue in quantum information theory  Vidal and Werner in \cite{PhysRevA.65.032314} proposed a computable measure termed as {\em entanglement negativity} which characterized the upper bound on the distillable entanglement whose non-convexity and monotonicity property was proved by Plenio in \cite{Plenio:2005cwa}. For a bipartite system $A \cup B$ in a pure and mixed state $\rho$, the entanglement negativity is defined as the logarithm of the trace norm of the partially transposed reduced density matrix as \cite{PhysRevA.65.032314}
\begin{equation}\label{ent_neg}
\mathcal{E} = \log \mathrm{Tr}|\rho^{T_B}|=\log(1+2 \sum_{\lambda_{i}<0}\left|\lambda_{i}\right|),
\end{equation}
where $\lambda_{i}$ are the negative eigenvalues of the matrix $\rho^{T_B}$. The partial transposed reduced density matrix $\rho^{T_B}$ of $\rho$ is obtained by exchanging the subsystem $B$'s qubit as  $|m~n\rangle\langle p~q|\to |m~q\rangle\langle m~n|$. Similar to the case of entanglement entropy, the authors in \cite{Calabrese:2012ew,Calabrese:2012nk,Calabrese:2014yza} described the computation of the entanglement negativity for bipartite mixed states in QFTs through a replica technique. For holographic characterization of the entanglement negativity of bipartite pure and mixed states in dual CFTs see \cite{Rangamani:2014ywa,Perlmutter:2015vma,Chaturvedi:2016rcn,Chaturvedi:2016opa,Malvimat:2017yaj,Chaturvedi:2016rft,Jain_2019hen,Jain:2017uhe,Jain:2017xsu,Jain:2018bai,Malvimat:2018txq,Malvimat:2018ood,Malvimat_2019,basak2020minimal,basak2021islands,basak2021page,afrasiar2021holographic,
basu2021entanglement,mondal2021holographic,Dong_2021,Kudler_Flam_2020,Kudler_Flam_2019} (and references therein).

On the other hand, investigation of entanglement in the non-inertial frame has witness surge in interest in the past few decades \cite{PhysRevLett.91.180404,Alsing_2004,PhysRevLett.95.120404,PhysRevA.74.032326,Hwang_2011,shamirzai2011tripartite,Alsing_2003,PhysRevA.81.032320,PhysRevA.80.042318,PhysRevA.82.042332,Hwang_2012}. Inertial observers are described by the Minkowski coordinates while Rindler coordinates are best suited for uniformly accelerated observers. A uniformly accelerated observer can not access the entire spacetime since it is constrained to move in one of the causally disconnected Rindler wedge. Therefore, the observer must trace over the inaccessible part thus losing information about the total state of the system. As a consequence the observer detects a thermal state which is known as Unruh effect \cite{PhysRevD.14.870}.

The authors in \cite{PhysRevLett.95.120404,PhysRevA.74.032326} studied the entanglement between two free modes of  bosonic and fermionic fields as seen by an inertial observer Alice detecting one of the  mode and a uniformly accelerating observer Rob detecting the other. Since the latter follows a hyperbolic trajectory in the Rindler coordinates which emerges as the near-horizon geometry of a Schwarzschild black hole, the above-mentioned scenario can be seen as entanglement between qubits located at horizon and that being sent away.  If the Hawking radiation would carry information, as a popular resolution to the information loss paradox, this entanglement could also contribute to the  time correlation between early and late radiation.  On the other hand, Hawking, Perry and Strominger noticed an infinite family of degenerate vacua associated with BMS supertranslation at null infinity.  They suggested an infinite numbers of soft hairs of black hole are responsible for storage of were-claimed-lost information \cite{Hawking:2016msc,Strominger:2017aeh,Perry:2020tts}.  However, the black hole loses its spherical isometry after generic supertranslation and a global Rindler coordinates becomes unavailable.  Nevertheless, it can be shown that locally the near-horizon geometry is no more than a shock wave background and we suggest to probe this anisotropy by placing the entangled qubit at different locations near horizon.

\begin{figure}[!]
\centering
\includegraphics[scale=0.4]{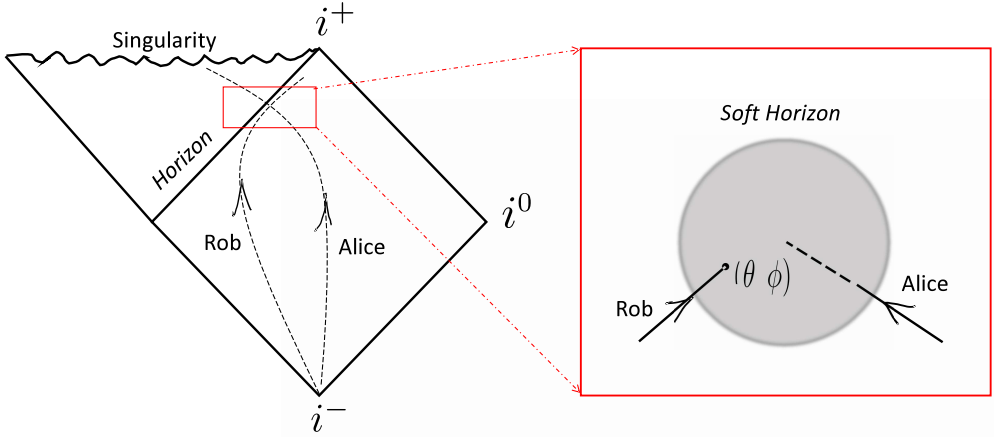}
\caption{\label{fig:alice_rob}A Gedanken scenario that Alice and Rob share an entangled qubit at past timelike infinity $i^-$.  Alice then free-falls towards the black hole while Rob hangs around at the horizon.  The degradation of entanglement might depend on Rob's angular location if the horizon were supertranslated by soft charges.}
\end{figure}

  To start with we consider a maximally entangled pure state (Bell state) in an inertial frame and study its entanglement at the  hairy horizon of a four-dimensional supertranslated Schwarzschild black hole. We consider two observers Alice and Rob, as pictured in the Fig. \ref{fig:alice_rob}, share an initially entangled state at the past timelike infinity $i^-$ before the formation of the black hole. Once the black hole is formed by the shock wave, they fall towards it and Rob eventually hovers outside the horizon whereas Alice continues its journey across the horizon. We investigate the entanglement between the two modes of a free bosonic and fermionic field as detected by Alice and Rob by computing  mutual information and entanglement negativity. We show that mutual information and entanglement negativity for both bosonic and fermionic case become angular dependent  by considering a specific of soft hair function. It has been discussed in \cite{Wen:2021ahw} that the supertranslated Schwarzschild metric induces uneven surface gravity at the hairy horizon, and we show that this causes different degradation of entanglement between Alice and Rob.  Furthermore, we show that for vanishing soft hair function, our results mimic the usual behavior as obtained in \cite{2008HR} for bosonic fields and in \cite{PhysRevA.74.032326} for fermionic fields.

  
%

The article is organized as follows. In section \ref{sfswb} we derive the shock wave background as the near-horizon geometry of Schwarzschild black hole under supertranslation. In section \ref{eswb}, we study entanglement between two relatively accelerating observers of free modes of  bosonic and fermionic fields at the soft hair horizon of a four-dimensional supertranslated Schwarzschild black hole by computing mutual information and entanglement negativity. In the final section \ref{sac}, we present a brief summary and our conclusions.

\section{Near hairy horizon geometry and shock wave background}\label{sfswb}

\begin{figure}[!]
\centering
\includegraphics[scale=0.4]{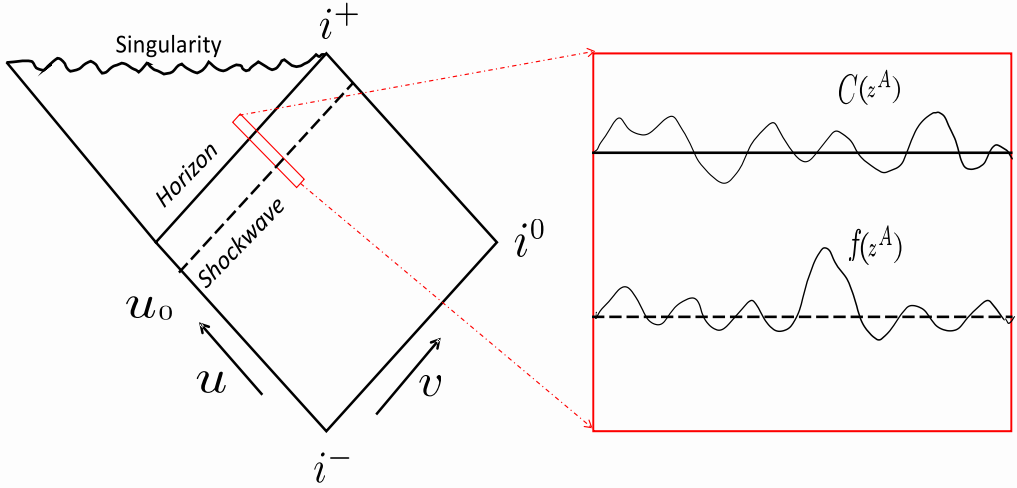}
\caption{\label{fig:shockwave}A black hole is formed by a shock wave where the soft hair function $C(z^A)$ is point-wisely mapped to the waveform factor $f(z^A)$ via the eq. (\ref{eqn:transform}).}
\end{figure}

We propose a $4$-dimensional spherical shock wave in Minkowski spacetime by the following metric, after the construction in  \cite{DRAY1985173,Comp_re_2019}
\begin{eqnarray}\label{Min_sw}
\mathrm{d} s^{2} &=&-\mathrm{d} u \mathrm{~d} v+f(z^A) \delta\left(u-u_{0}\right) \mathrm{d} u^{2}+ r^2\gamma_{AB}\mathrm{d} z^A \mathrm{d} z^B, \\
&=&-\mathrm{d} u \mathrm{~d} \hat{v}-\Theta\left(u-u_{0}\right) \partial_{B} f(z^A) \mathrm{d} u \mathrm{~d} z^{B}+ r^2\gamma_{AB}\mathrm{d} z^A \mathrm{d} z^B.
\end{eqnarray}
Here $u=t+r,~ v=t-r$ are the light-cone coordinates, and $z^A$ are transverse Fubuni-Study coordinates of $2$-sphere. The function $f(z^A)$ is non-negative \cite{Camanho_2016} which describes the shock wave form factor. The {\sl hat} coordinate is obtained by the {\sl supertranslation} shift
\begin{equation}
\hat{v}=v-\Theta\left(u-u_{0}\right) f(z^A),
\end{equation}
where $\Theta\left(u-u_{0}\right)$ is the heaviside step function. $T_{u u}=\delta\left(u-u_{0}\right) T(z^A)$ is the only non-zero component of the stress-tensor,  localized on the shock wave front $u=u_{0}$ which propagates on the negative $r$ direction.  At late stage, a Schwarzschild black hole is formed by collapsing this shock wave which is depicted in Fig. \ref{fig:shockwave}.  In the following, we will show this shock wave metric is related to the near-horizon metric of a $4$-dimensional supertranslated Schwarzschild black hole.

Recall under supertranslation, the $4$-dimensional hairy Schwarzschild black hole in isotropic coordinate $(t,\rho, z^A)$ becomes\cite{Compere:2016hzt}:
\begin{equation}\label{eqn:hair_metric}
ds^2=-\frac{(1-\frac{M}{2\rho_s})^2}{(1+\frac{M}{2\rho_s})^2}dt^2+(1+\frac{M}{2\rho_s})^4\bigg( d\rho^2+\big(\big((\rho-E)^2+U \big)\gamma_{AB} + (\rho-E)C_{AB}\big) dz^A dz^B \bigg),
\end{equation}
where scalars $E, U$ and tensor $C_{AB}$ are functions of $C(z^A)$, defined as follows:

\begin{eqnarray}
C_{AB}&=&-(2D_AD_B-\gamma_{AB}D^2)C,\nonumber\\
E&=&\frac{1}{2}D^2C+C-C_{0,0},\nonumber\\
U&=&\frac{1}{8}C_{AB}C^{AB}.
\end{eqnarray}
We remark that the {\sl supertranslated} radius  $\rho_s = \sqrt{(\rho-C-C_{0,0})^2+||{\cal D} C||^2}$.  Here $C_{0,0}$ refers to the zero mode in spherical harmonic expansion, which generates a time translation, and the square of norm $||{\cal D} C||^2 \equiv \gamma_{AB}D^AC D^BC$.  The metric reduces to the no-hair Schwarzschild solution for vanishing soft hair function $C$. We now take the near-horizon limit of hair metric eq. (\ref{eqn:hair_metric}), say $\rho_s \simeq \frac{M}{2}+Mx$ for $x \ll 1$\cite{Wen:2021ahw}:
\begin{equation}\label{eqn:near-horizon}
 ds^2 \simeq \underbrace{-x^2dt^2 + \kappa^{-2} dx^2}_{\text{Rindler space}} + \underbrace{\cdots}_{\text{Sphere part}},  
\end{equation}
where we identify the part in Rindler coordinates and obtain the angle-dependent surface gravity as
\begin{equation}\label{surfacegrav}
\kappa=\frac{\sqrt{M^2-4||{\cal D}C||^2}}{4M^2}.
\end{equation}
For small hair function $C$, the surface gravity in eq. \eqref{surfacegrav} maybe further expanded as
\begin{equation}\label{smallkappa}
\begin{aligned}
&\kappa{\simeq}(\kappa_0-\epsilon),\\
&\kappa_0=\frac{1}{4 M},~~~~~\epsilon=\frac{||{\cal D}C||^2}{2 M^3}.
\end{aligned}
\end{equation}
We can identify $\kappa_0$  as the surface gravity without any soft hair, and regard $\epsilon$ as a correction to the surface gravity $\kappa$. As an explicit example, we will adopt the following soft hair function \cite{Comp_re_2016,Lin_2020}
\begin{equation}
\begin{aligned}
C= M\epsilon^{\prime} Y_{2,0}(\theta, \phi)= M\epsilon^{\prime} \sqrt{\frac{5}{16 \pi}}\left(3 \cos ^{2} \theta-1\right),
\end{aligned}
\end{equation}
for sufficiently small $\epsilon^{\prime}$ and $(\theta,\phi)$ in $Y_{2,0}(\theta, \phi)$ are the angular coordinates on a unit $2$-sphere. Then $\epsilon$ maybe computed from eq. \eqref{smallkappa}, which is given as
\begin{equation}\label{epsilon}
\epsilon=\frac{45 \sin^2{2\theta}{\epsilon^{\prime}}^2}{32 M}.
\end{equation}
Note that $\epsilon^{\prime}\leq .42$ in eq. \eqref{epsilon} to ensure the positivity of the surface gravity $\kappa$ in eq.  \eqref{smallkappa}. Thus, the angular dependence of the  surface gravity in eq. \eqref{surfacegrav} is manifest through the correction eq. \eqref{epsilon}.

Recall that we are interested in  investigating the entanglement between the free falling observer, Alice and the accelerating observer, Rob at the soft hair horizon. For simplicity, we assume that Rob approaches the hairy black hole at a constant angle, therefore the hair function remains constant.  At each angular patch $U_i(z^A_i)$ or equivalently $U_i(\theta_i,\phi_i)$, the Rindler part of metric eq.  (\ref{eqn:near-horizon}) can be brought into the shock wave form eq. (\ref{Min_sw}) under the transformation\footnote{For small enough patch, one can also assume the wave form $f(z^A)$ remains constant.}:
\begin{eqnarray}\label{eqn:transform}
 u&=&-\kappa_i^{-1}x e^{-\kappa_i t},\nonumber\\
 \hat{v}&=&\kappa_i^{-1}x e^{\kappa_i t},
\end{eqnarray}
where the constant $\kappa_i = \kappa|_{z^A_i}$ .  By studying the transition between two neighboring patches, one can further find out the differential relation between hair function $C(z^A)$ and wave form factor $f(z^A)$:
\begin{equation}
      \partial_{z^A}||{\cal D}C(z^A)|| \propto \partial_{z^A}f(z^A)   
\end{equation}
We leave the construction detail in the appendix.  In summary, Rob near the hairy horizon can be regarded as a Rindler observer in a shock wave background.  In the next section, we will take advantage of this equivalence and compute his entanglement with free-falling observer Alice.

\section{Entanglement at the soft-hair horizon of a supertranslated Schwarzschild black hole}\label{eswb}
We now proceed to study the entanglement between two observers
Alice  and Rob of a $4$-dimensional supertranslated Schwarzschild black hole background by computing mutual information and entanglement negativity. As mentioned before, we consider Alice and Rob share an entangled state (Bell state) at the same initial point in flat Minkowski spacetime before the formation of the black hole. Once the black hole is formed by the shock wave, Alice free falls into the black hole, while Rob accelerates following a hyperbolic trajectory near the soft hair horizon of a supertranslated Schwarzschild black hole. The maximally entangled Bell state is given as 
\begin{equation}\label{bellstate}
\left|\psi\right\rangle=\frac{1}{\sqrt{2}}\left(\left|0\right\rangle_{\mathrm{A}}\left|0\right\rangle_{\mathrm{R}}+\left|1\right\rangle_{\mathrm{A}}\left|1\right\rangle_{\mathrm{R}}\right),
\end{equation}
Note that we assume Alice has a detector which only detects $\left|n\right\rangle_{\mathrm{A}}$ mode while Rob detects only mode $\left|n\right\rangle_{\mathrm{R}}$. The states corresponding to Rob $\left|n\right\rangle_{\mathrm{R}}$ must be expressed in terms of the black hole coordinates \cite{2008HR} in order to express what Rob observes in the curved spacetime near the soft hair horizon of the supertranslated Schwarzschild black hole.
\subsection{Entanglement for bosonic field}
\begin{figure}[!]
\centering
\begin{subfigure}{.5\textwidth}
  \centering
  \includegraphics[width=1\linewidth]{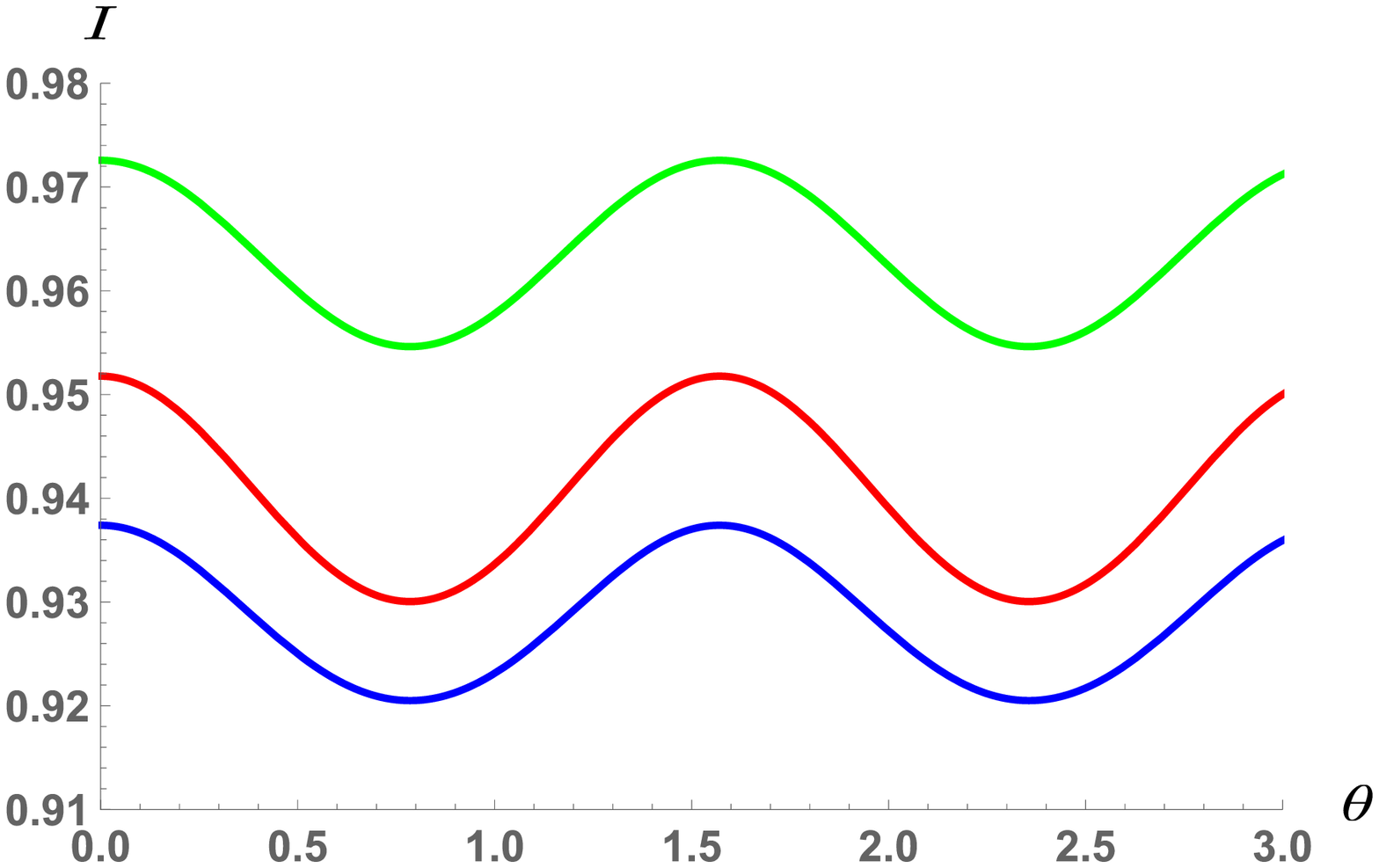}
  \caption{}
  \label{mibfwh}
\end{subfigure}\hfill%
\begin{subfigure}{.5\textwidth}
  \centering
  \includegraphics[width=1\linewidth]{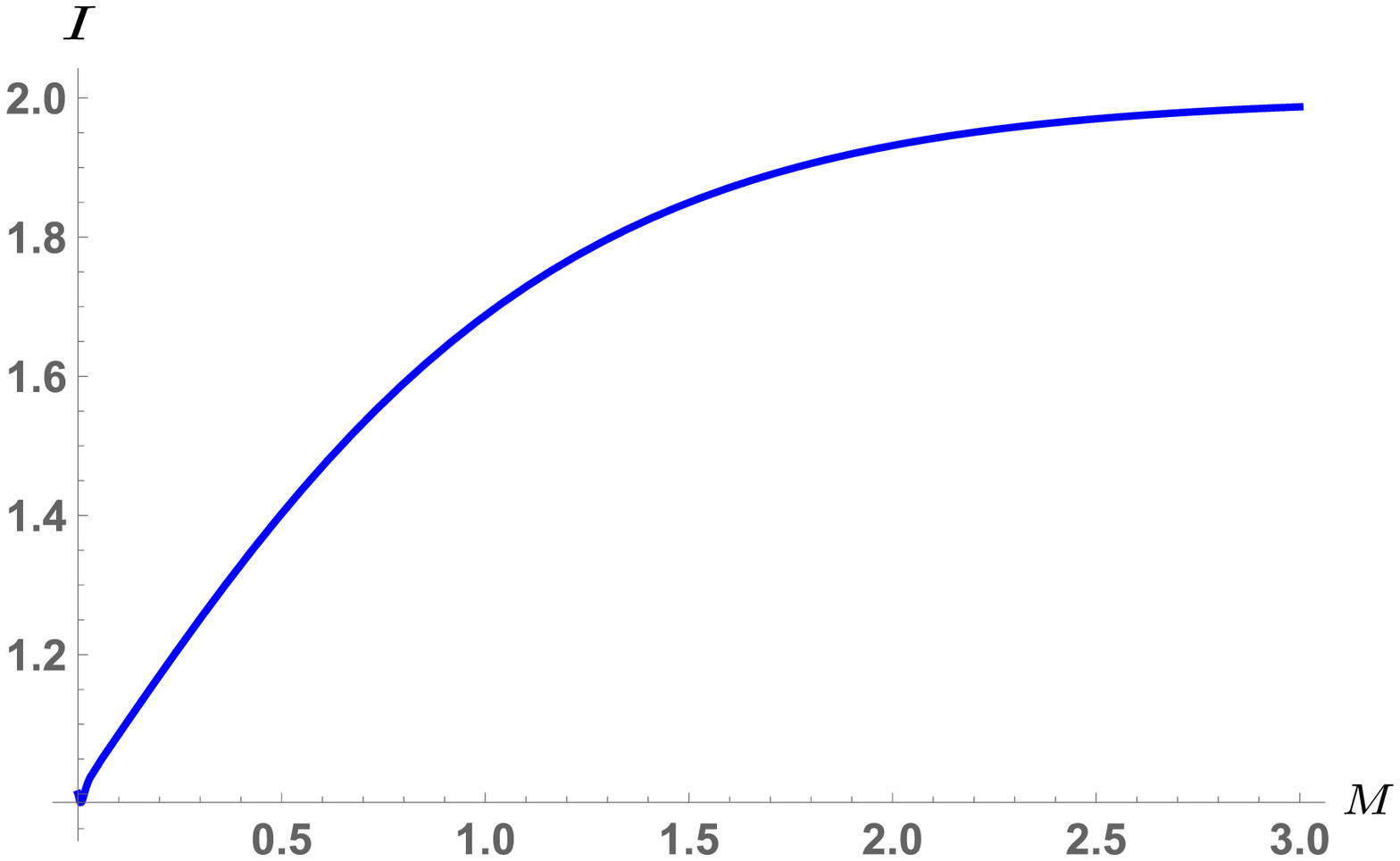}
  \caption{}
  \label{mibfnh}
\end{subfigure}
\caption{(a) Mutual information $I(A:R)$ is plotted as a function of $\theta$ with a fixed field mode frequency $\omega=1/4 \pi$ and $\epsilon^{\prime}=0.3$ for different black hole mass $M=.03$ (green curve), $M=.05$ (red curve) and $M=.07$ (blue curve). (b) Mutual information $I(A:R)$ is plotted as a function of  black bole mass $M$ for vanishing soft hair function $C=0$. }
\label{}
\end{figure}

In this section we compute mutual information and entanglement negativity for two free bosonic modes between Alice and Rob. The vacuum $\left|0\right\rangle_{\mathrm{R}}$ and the first-excited state $\left|1\right\rangle_{\mathrm{R}}$ of Rob maybe expressed in terms of two-mode squeezed state as \cite{2008HR}
\begin{equation}\label{grstfirst1}
\begin{aligned}
 |0\rangle_{R}&=\sqrt{1-e^{-2 \pi \omega / \kappa}} \sum_{n=0}^{\infty} e^{-n \pi \omega / \kappa}|n\rangle_{\text {in }} \otimes|n\rangle_{\text {out }},\\
|1\rangle_{R}&=\left(1-e^{-2 \pi \omega / \kappa}\right) \sum_{n=0}^{\infty} \sqrt{n+1} e^{-n \pi \omega / \kappa}|n\rangle_{i n} \otimes|n+1\rangle_{o u t},
\end{aligned}
\end{equation}
where $\left\{|n\rangle_{\text {in }}\right\}$ and $\left\{|n\rangle_{\text {out }}\right\}$ are the orthonormal bases for the inside and outside region of the event horizon respectively. Employing eq. \eqref{smallkappa}, the above eq. \eqref{grstfirst1} maybe expressed as follows
\begin{equation}\label{grstfirst2}
\begin{aligned}
 |0\rangle_{R}&=\sqrt{1-(1-2 \alpha ) e^{-2 x}} \sum_{n=0}^{\infty}e^{-n x} (1 -  n \alpha)|n\rangle_{\text {in }} \otimes|n\rangle_{\text {out }},\\
|1\rangle_{R}&=\sqrt{1-(1-2 \alpha ) e^{-2 x}} \sum_{n=0}^{\infty} \sqrt{n+1} e^{-n x} (1 -  n \alpha)|n\rangle_{i n} \otimes|n+1\rangle_{o u t},
\end{aligned}
\end{equation}
where
\begin{equation}\label{xnadalpha}
x=4 \pi \omega M,~~~\alpha=\frac{45 \pi \omega M{\epsilon^{\prime}}^2 \sin^2 2\theta}{2} .
\end{equation}
Note that henceforth we will denote $|n\rangle_{i n} \otimes|m\rangle_{o u t}$ simply as $|nm\rangle$. Furthermore, using  eq. \eqref{grstfirst2},  eq. \eqref{bellstate} maybe expressed in terms of Minkowski modes for Alice and black hole modes in the inside and outside regions of the soft hair horizon for Rob. Since Rob is causally disconnected from the inside region of the black hole, it is required to trace over the states from this part, and consequently we obtained the joint mixed density matrix of Alice and Rob which is given as  \cite{2008HR} 
\begin{equation}\label{ARdm}
\begin{aligned}
\rho_{A R} &=\big(1-(1-2 \alpha ) e^{-2 x}\big) \sum_{n=0}^{\infty}(1-2 \alpha  n) e^{-2 n x}~\rho_{n},
\end{aligned}
\end{equation}
where
\begin{equation}
\begin{aligned}
\rho_{n} =&\frac{1}{2}\Big(|0 n\rangle\langle 0 n|+\sqrt{(n+1)\big(1-(1-2 \alpha ) e^{-2 x}\big)}\left(| 0 n\rangle\langle 1 n+1|+|1 n+1\rangle\langle 0 n|\right)\\
&+{(n+1)} \big(1-(1-2 \alpha ) e^{-2 x}\big)| 1 n+1\rangle\langle 1 n+1|\Big).
\end{aligned}
\end{equation}
It is important to notice that the density matrix a $\rho_{AR}$ has $2\times 2$ block structure in the basis $\{|0~ n\rangle,|1 ~n+1\rangle\}_{n=0}^{\infty}$, i.e  the elements of the $(n,n+1)$ block of the matrix $\rho_{AR}$ is 
\begin{equation}\label{rarbd1}
\frac{1}{2} (1-2 \alpha  n) e^{-2 n x} \big(1-(1-2 \alpha ) e^{-2 x}\big)\left(\begin{array}{ll}
1 &  \sqrt{(n+1)\big(1-(1-2 \alpha ) e^{-2 x}\big)} \\
  \sqrt{(n+1)\big(1-(1-2 \alpha ) e^{-2 x}\big)}& (n+1)\big(1-(1-2 \alpha ) e^{-2 x}\big)
\end{array}\right).
\end{equation}
The entanglement entropy of this joint state $\rho_{AR}$ maybe computed by finding the eigenvalues of the above matrix, which is given as

\begin{equation}\label{entAR}
\begin{aligned}
S\left(\rho_{A R}\right)=&-\frac{1}{2}\sum_{n=0}^{\infty} (1-2 \alpha  n) e^{-2 n x} \big(1-(1-2 \alpha ) e^{-2 x}\big)\left[1+(n+1)\big(1-(1-2 \alpha ) e^{-2 x}\big)\right] \\
& \times \log _{2} \frac{1}{2}(1-2 \alpha  n) e^{-2 n x} \big(1-(1-2 \alpha ) e^{-2 x}\big)\left[1+(n+1)(1-(1-2 \alpha ) e^{-2 x})\right].
\end{aligned}
\end{equation}
Rob's density matrix is obtained by tracing over Alice's state as $\rho_{R}=\text{Tr}_A\left(\rho_{A R}\right)$, which reads
\begin{equation}
\rho_{R}=\frac{1}{2}(1-(1-2 \alpha ) e^{-2 x}) \sum_{n=0}^{\infty}(1-2 \alpha  n) e^{-2 n x}\left[|n\rangle\langle n|+(n+1)\big(1-(1-2 \alpha ) e^{-2 x}\big)| n+1\rangle\langle n+1|\right],
\end{equation}
and Rob's entropy is 
\begin{equation}\label{entR}
\begin{aligned}
S\left(\rho_{R}\right)=&-\frac{1}{2}\sum_{n=0}^{\infty} (1-2 \alpha  n) e^{-2 n x}\big(1-(1-2 \alpha ) e^{-2 x}\big)\left[1+n\big((1+2 \alpha) e^{2 x}-1\big)\right] \\
& \times \log _{2} \frac{1}{2} (1-2 \alpha  n) e^{-2 n x} \big(1-(1-2 \alpha ) e^{-2 x}\big)\left[1+n\big((1+2 \alpha) e^{2 x}-1\big)\right].
\end{aligned}
\end{equation}
Similarly, Alice's density matrix is obtained by tracing over Rob's state as $\rho_{A}=\text{Tr}_R\left(\rho_{A R}\right)$, which is given as
\begin{equation}
\rho_{A}=\frac{1}{2}\left(|0\rangle\left\langle 0\left|+\right| 1\right\rangle\langle 1|\right),
\end{equation}
and the associated entanglement entropy is $S\left(\rho_{R}\right)=1$. The mutual information $I(A:R)$ maybe computed between Alice and Rob by using eq. \eqref{muinfo} which is plotted as function of $\theta$ as shown in Fig. \ref{mibfwh}. Note that in Fig. \ref{mibfnh} we plot mutual information $I(A:R)$ as a function of black hole mass (or inverse Hawking temperature) for vanishing soft hair function (i.e $C=0$), which reproduces the  mutual information for Alice and  Rob as obtained for the case of ordinary Schwarzschild black hole in \cite{2008HR}.

\begin{figure}
\centering
\begin{subfigure}{.5\textwidth}
  \centering
  \includegraphics[width=1\linewidth]{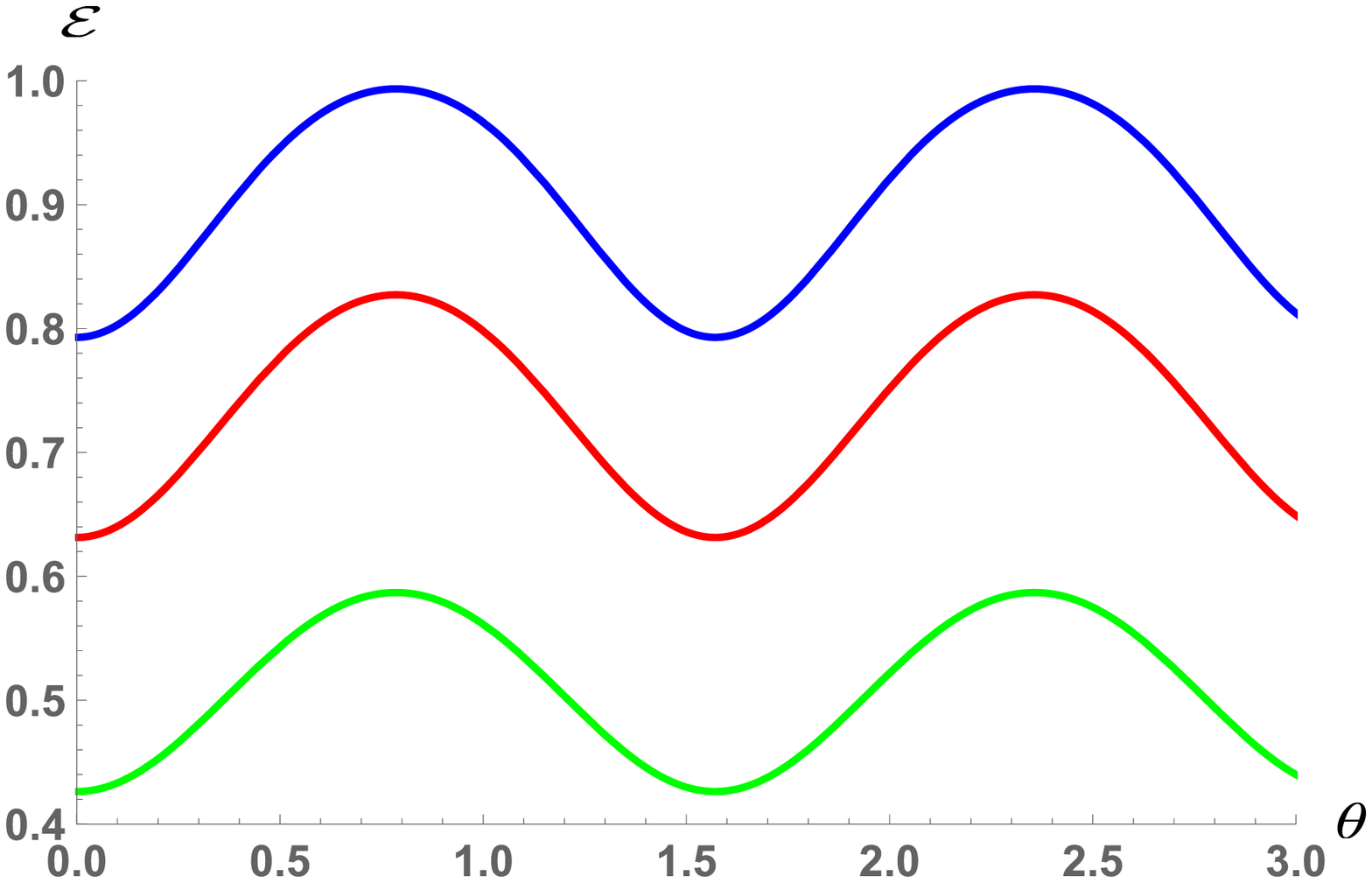}
  \caption{}
  \label{enbfwh}
\end{subfigure}%
\begin{subfigure}{.5\textwidth}
  \centering
  \includegraphics[width=1\linewidth]{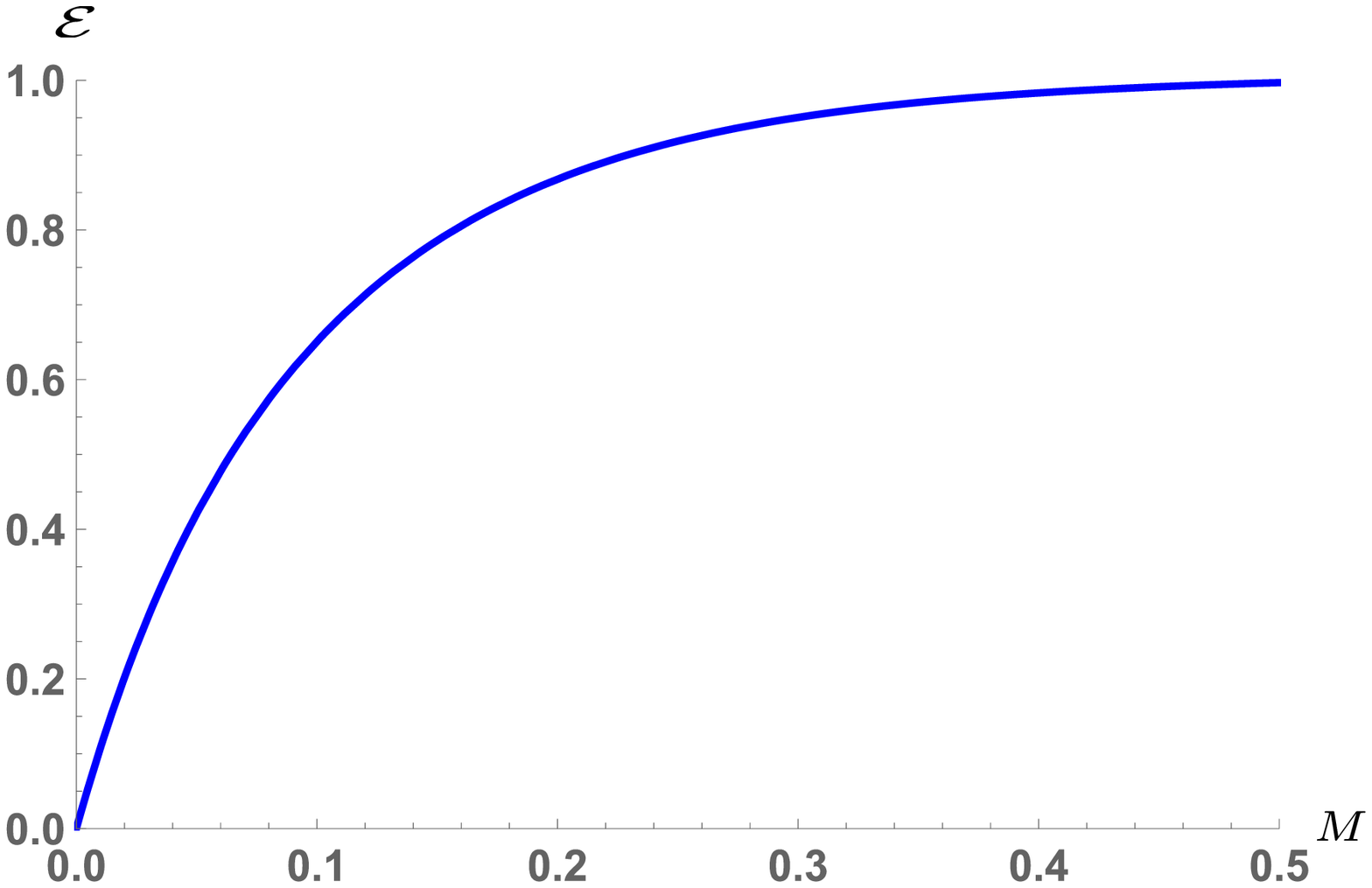}
  \caption{}
  \label{enbfnh}
\end{subfigure}
\caption{(a) Entanglement negativity $\mathcal{E}$ is plotted as a function of $\theta$ with a fixed field mode frequency $\omega=1/4 \pi$ and $\epsilon^{\prime}=0.3$ for different black hole mass $M=.03$ (green curve), $M=.05$ (red curve) and $M=.07$ (blue curve). (b) Entanglement negativity $\mathcal{E}$ is plotted as a function of  black bole mass $M$ for vanishing soft hair function $C=0$.}
\label{}
\end{figure}

To compute entanglement negativity between Alice and Rob, we need to partially transpose the bipartite density matrix $\rho_{AR}$ in eq. \eqref{ARdm}. The partial transpose density matrix $\rho_{AR}^{T_A}$ is obtained by  interchanging the Alice's qubit (i.e $|m~n\rangle\langle p~q|\to |p~n\rangle\langle m~q|$) which has a $2\times 2$ block structure in the basis $\{|0~ n+1\rangle,|1 ~n\rangle\}_{n=0}^{\infty}$, which is given as 
\begin{equation}\label{rarbd11}
\frac{1}{2} \left(1-2 \alpha  n) e^{-2 n x} \left(1-(1-2 \alpha \right) e^{-2 x}\right)\left(\begin{array}{ll}
n \left((2 \alpha +1) e^{2 x}-1)\right) &  \sqrt{(n+1) \left(1-(1-2 \alpha ) e^{-2 x}\right)}\\
 \sqrt{(n+1) \left(1-(1-2 \alpha ) e^{-2 x}\right)}& (1-2 \alpha ) e^{-2 x}
\end{array}\right).
\end{equation}
The eigenvalues of the above matrix are as follows
\begin{equation}\label{eigenv}
\begin{aligned}
\lambda_{1}^{n}&=-\frac{1}{4} (2 \alpha  n-1) e^{-2 (n+2) x} \left(2 \alpha +e^{2 x}-1\right)\Bigg(\xi_n+\sqrt{4 e^{2 x} \Big(2 \alpha +e^{2 x} \left(4 \alpha ^2 n+1\right)-1\Big)+\xi_n^2}\Bigg),\\
   \lambda_{2}^{n}&=-\frac{1}{4} (2 \alpha  n-1) e^{-2 (n+2) x} \left(2 \alpha +e^{2 x}-1\right)\Bigg(\xi_n-\sqrt{4 e^{2 x} \Big(2 \alpha +e^{2 x} \left(4 \alpha ^2 n+1\right)-1\Big)+\xi_n^2}\Bigg),
   \end{aligned}
\end{equation}
where
\begin{equation}
\xi_n=1-2 \alpha +(2 \alpha +1) n e^{4 x}-n e^{2 x}.
\end{equation}
The eigenvalues $\lambda_2^n$  are negative for all possible values of $x$ and $\alpha$, and the associated entanglement negativity is obtained from eq. \eqref{ent_neg} which is given as follows 
\begin{equation}\label{entforbosonfield}
\begin{aligned}
\mathcal{E} &= \log(1+2 \sum_{n=0}^\infty|\lambda^n_{2}|).
\end{aligned}
\end{equation}
The entanglement negativity depends on the angular coordinate $\theta$ through the function $\alpha$ which we plot in Fig. \ref{enbfwh}. For vanishing soft hair function $C=0$, the eq. \eqref{entforbosonfield} reproduces the entanglement negativity  between Alice and Rob as that obtained for the case of an ordinary Schwarzschild black hole in \cite{2008HR}, which we depict in Fig. \ref{enbfnh} as a function of black hole mass $M$.

\subsection{Entanglement for fermionic field}
\begin{figure}
\centering
\begin{subfigure}{.5\textwidth}
  \centering
  \includegraphics[width=1\linewidth]{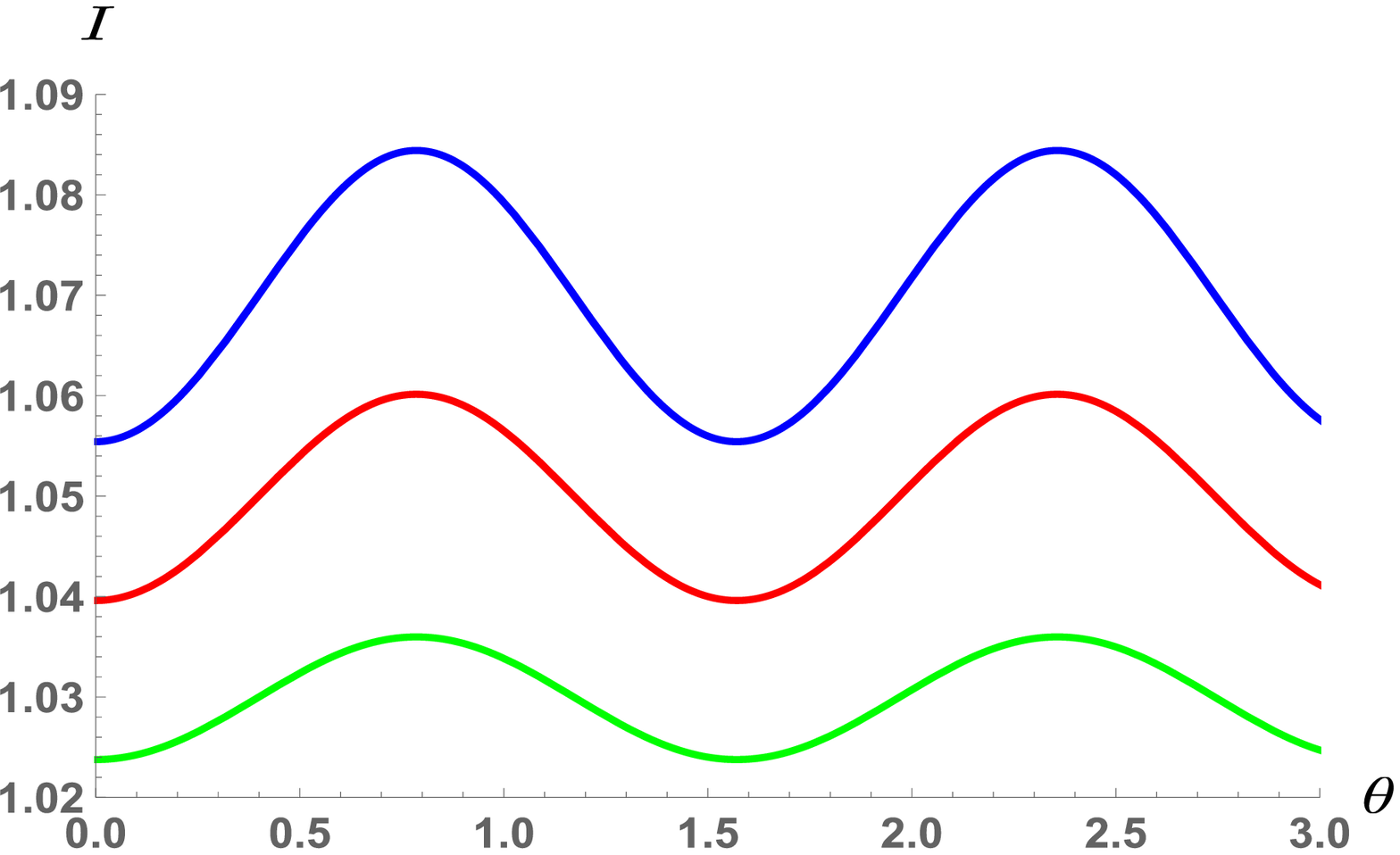}
  \caption{}
  \label{miffwh}
\end{subfigure}%
\begin{subfigure}{.5\textwidth}
  \centering
  \includegraphics[width=1\linewidth]{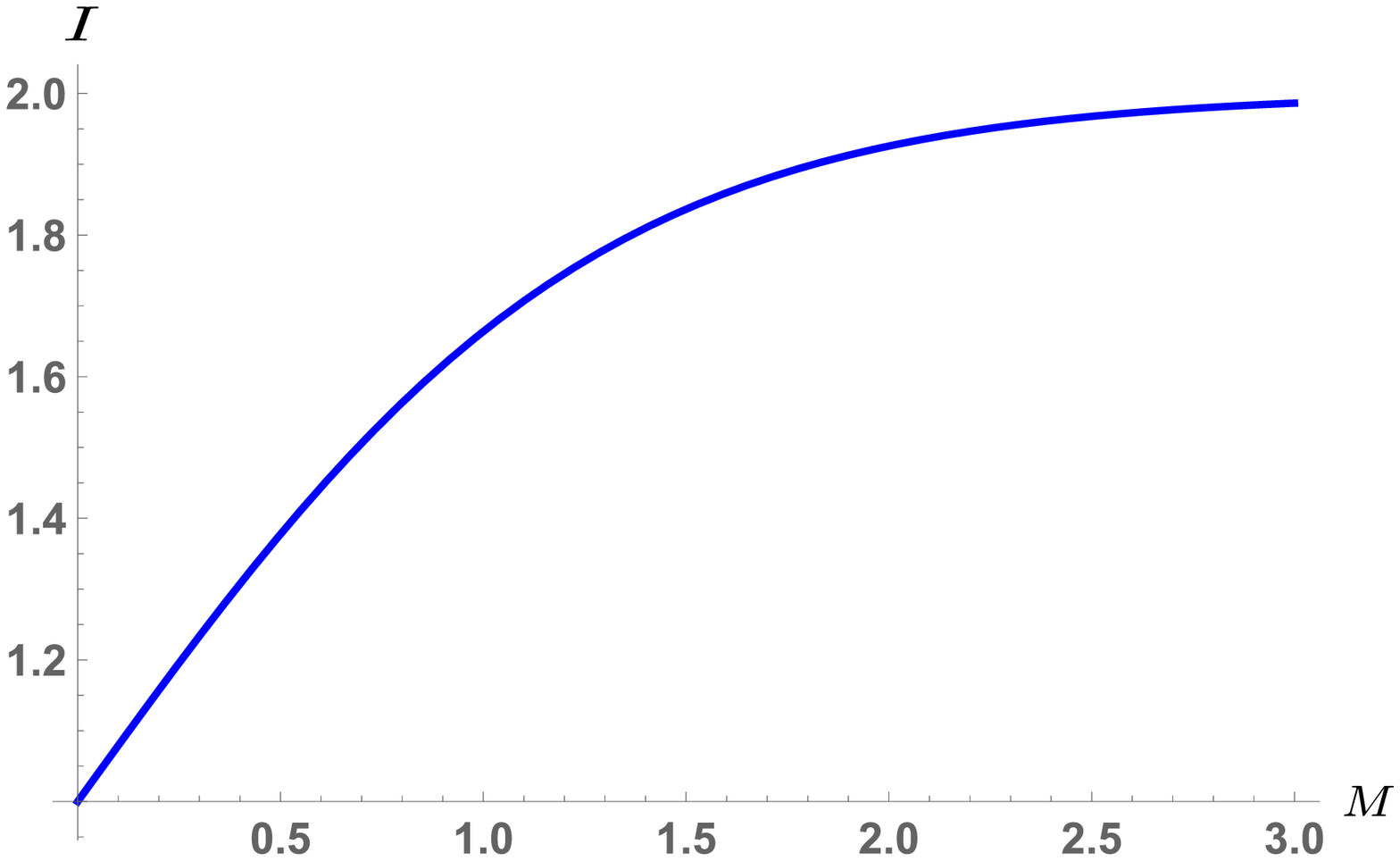}
  \caption{}
  \label{miffnh}
\end{subfigure}
\caption{(a) Mutual information $I(A:R)$ is plotted as a function of $\theta$ with a fixed field mode frequency $\omega=1/4 \pi$ and $\epsilon^{\prime}=0.3$ for different black hole mass $M=.03$ (green curve), $M=.05$ (red curve) and $M=.07$ (blue curve). (b) Mutual information $I(A:R)$ is plotted as a function of  black bole mass $M$ for vanishing soft hair function $C=0$.}
\label{}
\end{figure}

In this section we proceed to compute mutual information and entanglement negativity between two free fermionic modes one is shared by Alice and the other one by Rob near soft hair horizon of a four dimensional supertranslated Schwarzschild black hole. The Kruskal vacuum state $\left|0\right\rangle_{\mathrm{R}}$ and the first-excited state $\left|1\right\rangle_{\mathrm{R}}$ of Rob maybe expressed in terms of two-mode squeezed state as 
\cite{PhysRevA.74.032326}
\begin{equation}\label{grstfirstfer2}
\begin{aligned}
 |0\rangle_{R}&=\cos \frac{\pi \omega}{\kappa} |0\rangle_{\text {out }}+ \sin\frac{\pi \omega}{\kappa} |1\rangle_{\text {in }} |1\rangle_{\text {out }},\\
|1\rangle_{R}&=|1\rangle_{\text {in }} |0\rangle_{\text {out }},
\end{aligned}
\end{equation}
where $\kappa$ is the surface gravity at the soft hair horizon. Employing eq. \eqref{smallkappa}, $\cos \frac{\pi \omega}{\kappa}$ and $\sin\frac{\pi \omega}{\kappa}$ in the above equation may be expanded as follows 
\begin{equation}\label{grstfirstfer21}
\begin{aligned}
 \cos \frac{\pi \omega}{\kappa}&=\frac{1}{\sqrt{1+(1-2\alpha)e^{-2x}}},\\
\sin\frac{\pi \omega}{\kappa}&=\sqrt{\frac{(1-2\alpha)e^{-2x}}{1+(1-2\alpha)e^{-2x}}},
\end{aligned}
\end{equation}
where $\alpha$ and $x$ are given in eq. \eqref{xnadalpha}. The Bell state in eq. \eqref{bellstate} may then be expressed by using eqs. \eqref{grstfirstfer2} and \eqref{grstfirstfer21} as
\begin{equation}
|\psi\rangle=\frac{1}{\sqrt{2}}\Big(\frac{1}{\sqrt{1+(1-2\alpha)e^{-2x}}}|000\rangle+\sqrt{\frac{(1-2\alpha)e^{-2x}}{1+(1-2\alpha)e^{-2x}}}|011\rangle+|110\rangle\Big).
\end{equation}
The total density matrix of the system is $\rho=|\psi\rangle\langle \psi |$. The reduced density matrix for Alice-Rob $\rho_{AR}$ is given by
\begin{equation}\label{fpee1}
\begin{aligned}
\rho_{AR}=&\frac{1}{2}\Big(\frac{1}{{1+(1-2\alpha)e^{-2x}}} |00\rangle\langle 00 |+  \frac{1}{\sqrt{1+(1-2\alpha)e^{-2x}}} |00\rangle\langle 11 |\\
&+ {\frac{(1-2\alpha)e^{-2x}}{1+(1-2\alpha)e^{-2x}}} |01\rangle\langle 01 | + \frac{1}{\sqrt{1+(1-2\alpha)e^{-2x}}} |11\rangle\langle 00 |+ |11\rangle\langle 11 |\Big),
\end{aligned}
\end{equation}
which in the matrix form maybe expressed as 
\begin{equation}
\frac{1}{2}\begin{pmatrix}
\frac{1}{{1+(1-2\alpha)e^{-2x}}} &  0&0&\frac{1}{\sqrt{1+(1-2\alpha)e^{-2x}}} \\
  0& {\frac{(1-2\alpha)e^{-2x}}{1+(1-2\alpha)e^{-2x}}} &0&0\\
  0&0&0&0\\
  \frac{1}{\sqrt{1+(1-2\alpha)e^{-2x}}} \big)&0&0&1
\end{pmatrix},
\end{equation}
in the basis $\{|00\rangle,|01\rangle,|10\rangle,|11\rangle\}$ where $|a b\rangle=|a\rangle_{A}|b\rangle_{R}$. The associated entanglement entropy of Alice and Rob is 
\begin{equation}\label{entofAR}
\begin{aligned}
S(\rho_{AR})=&-\frac{2+(1-2 \alpha ) e^{-2 x}}{2\big(1+(1-2\alpha)e^{-2x}\big)}\log\Big(\frac{2+(1-2 \alpha ) e^{-2 x}}{2\big(1+(1-2\alpha)e^{-2x}\big)}\Big)\\
&-\frac{(1-2 \alpha ) e^{-2 x}}{2\big(1+(1-2\alpha)e^{-2x}\big)}\log\Big(\frac{(1-2 \alpha ) e^{-2 x}}{2\big(1+(1-2\alpha)e^{-2x}\big)}\Big).
\end{aligned}
\end{equation}
The reduced density matrices of Rob $\rho_R$ is
\begin{equation}\label{fpee1}
\begin{aligned}
\rho_{R}&=\frac{1}{2}\Big(\frac{1}{{1+(1-2\alpha)e^{-2x}}}|0\rangle\langle 0 |+ 1+\frac{(1-2\alpha)e^{-2x}}{1+(1-2\alpha)e^{-2x}}|1\rangle\langle 1 |\Big),
\end{aligned}
\end{equation}
which has the following matrix representation in the basis $\{|00\rangle,|11\rangle\}$
\begin{equation}
\rho_{R}=\frac{1}{2}\begin{pmatrix}
\frac{1}{{1+(1-2\alpha)e^{-2x}}} &  0 \\
  0&1+\frac{(1-2\alpha)e^{-2x}}{1+(1-2\alpha)e^{-2x}}
\end{pmatrix}.
\end{equation}
The entanglement entropy of Rob is
\begin{equation}\label{entofR}
\begin{aligned}
S(\rho_{R})=&-\frac{1}{2\big(1+(1-2 \alpha ) e^{-2 x}\big)}\log\Big(\frac{1}{2\big(1+(1-2 \alpha ) e^{-2 x}\big)}\Big)\\
&-\Big(1-\frac{1}{2\big(1+(1-2 \alpha ) e^{-2 x}\big)}\Big)\log\Big(1-\frac{1}{2\big(1+(1-2 \alpha ) e^{-2 x}\big)}\Big).
\end{aligned}
\end{equation}
Similarly, the density matrix of Alice $\rho_A$ is
\begin{equation}\label{fpee2}
\begin{aligned}
\rho_{A}&=\frac{1}{2}\Big( |0\rangle\langle 0 |+  |1\rangle\langle 1 |\Big),\\
\end{aligned}
\end{equation}
which maybe written in the matrix form in the basis $\{|00\rangle,|11\rangle\}$ as
\begin{equation}\label{entalice}
\rho_{A}=\frac{1}{2}\begin{pmatrix}
1 &  0\\
  0&1
\end{pmatrix},
\end{equation}
with entanglement entropy $S(\rho_{A})=1$. The mutual information $I(A:R)$  eq. \eqref{muinfo}  between Alice and Rob maybe obtained by employing eqs. \eqref{entofAR} and  \eqref{entofR}, which is plotted as a function of 
the angular coordinate $\theta$ for different black hole mass $M$ in Fig. \ref{miffwh}. Note that for $\alpha=0$ (i.e vanishing soft hair function $C=0$), the mutual information becomes equal to that of a no hair Schwarzschild black hole case as obtained in \cite{PhysRevA.74.032326} which we plot as a function mass $M$ in Fig. \ref{miffnh}.

%

%

\begin{figure}
\centering
\begin{subfigure}{.5\textwidth}
  \centering
  \includegraphics[width=1\linewidth]{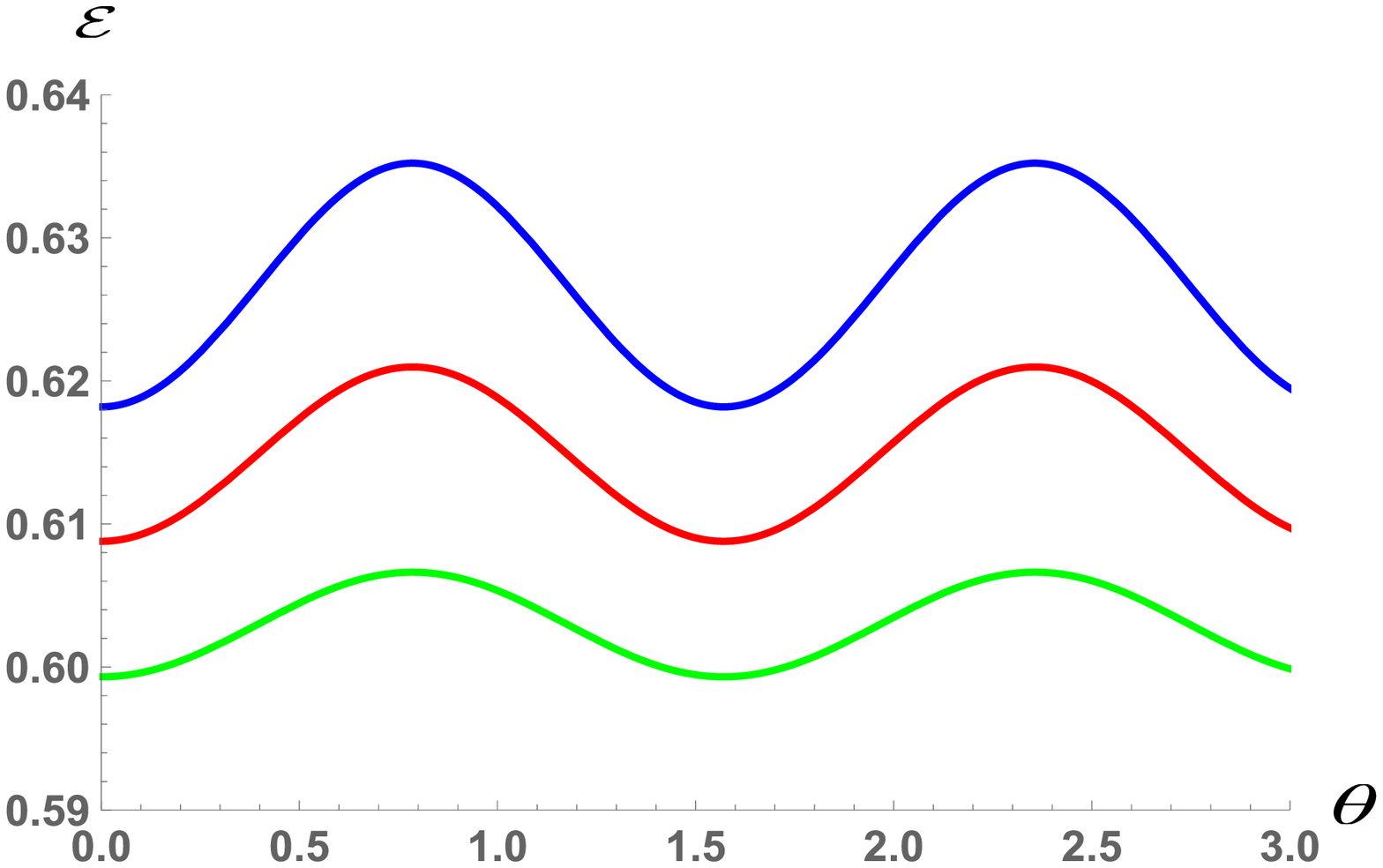}
  \caption{}
  \label{enffwh}
\end{subfigure}
\begin{subfigure}{.5\textwidth}
  \centering
  \includegraphics[width=1\linewidth]{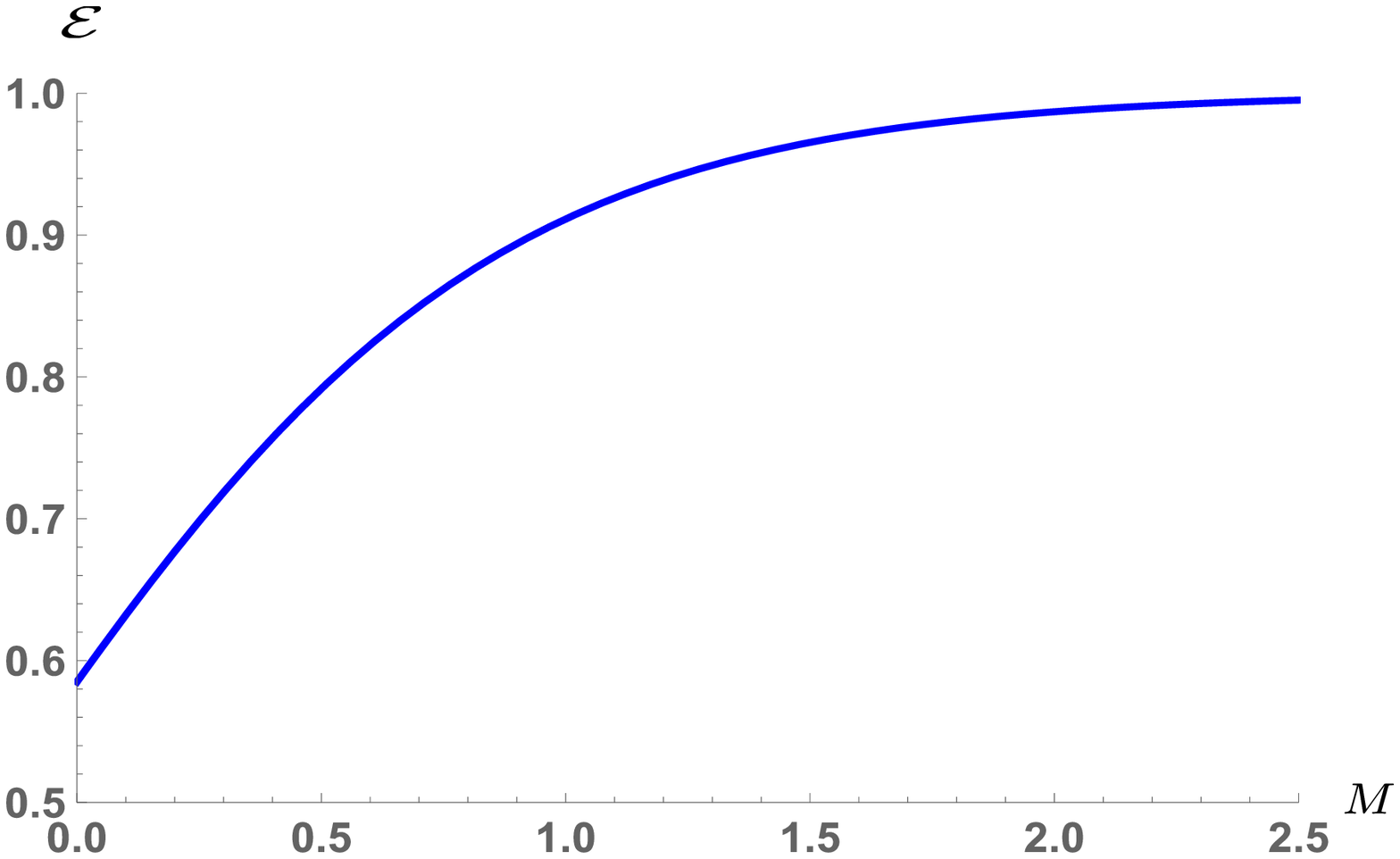}
  \caption{}
  \label{enffnh}
\end{subfigure}
\caption{(a) Entanglement negativity $\mathcal{E}$ is plotted as a function of $\theta$ with a fixed field mode frequency $\omega=1/4 \pi$ and $\epsilon^{\prime}=0.3$ for different black hole mass $M=.03$ (green curve), $M=.05$ (red curve) and $M=.07$ (blue curve). (b) Entanglement negativity $\mathcal{E}$ is plotted as a function of  black bole mass $M$ for vanishing soft hair function $C=0$.}
\label{}
\end{figure}

To compute entanglement negativity, it is required to obtain the partial transpose density matrix of $\rho_{AR}$ by interchanging the Alice's qubit which is given as

\begin{equation}
\rho_{AR}^{T_A}=\frac{1}{2}\begin{pmatrix}
\frac{1}{1+(1-2\alpha)e^{-2x} }&  0&0&0 \\
  0& {\frac{(1-2\alpha)e^{-2x}}{1+(1-2\alpha)e^{-2x}}} &\frac{1}{\sqrt{1+(1-2\alpha)e^{-2x}}} &0\\
  0&\frac{1}{\sqrt{1+(1-2\alpha)e^{-2x}}} &0&0\\
  0&0&0&1
\end{pmatrix}.
\end{equation}
The eigenvalues of the above matrix are as follows
\begin{equation}\label{eigenvaluesptdm}
\begin{aligned}
\lambda_1=\frac{1}{2},~
\lambda_2=\frac{1}{2},~
\lambda_3=\frac{1}{2}\frac{1}{1+(1-2\alpha)e^{-2x} },~
\lambda_4=-\frac{1}{2}\frac{1}{1+(1-2\alpha)e^{-2x} }.
\end{aligned}
\end{equation}
The eigenvalue $\lambda_4$ is negative and the
entanglement negativity is obtained by using eq. \eqref{ent_neg}, which is given as
\begin{equation}
\begin{aligned}
\mathcal{E} =& \log(1+2 |\lambda_4|),\\
=&\log\Big(\frac{2+(1-2 \alpha ) e^{-2 x}}{1+(1-2 \alpha ) e^{-2 x}}\Big).
\end{aligned}
\end{equation}
The entanglement negativity in the above equation depends of the hair function through $\alpha$. In Fig. \ref{enffwh}, we plot entanglement negativity as a function of angle $\theta$ for various black hole mass. Vanishing value of $\alpha~(\textrm{i.e}~ C=0)$ represents the no hair black hole scenario and the  corresponding entanglement negativity becomes equal to that of an ordinary Schwarzschild black hole case as computed in \cite{PhysRevA.74.032326} which we plot as a function of black hole mass $M$ in Fig. \ref{enffnh}.

\section{Summary and Conclusion}\label{sac}
In this article we have studied entanglement between the two modes of a free bosonic and fermionic field as detected by two relatively accelerating observers Alice who is free falling and Rob who is hanging near the soft hair horizon of a four-dimensional supertranslated Schwarzschild black hole. The entanglement between these two observers was measured by investigating mutual information and entanglement negativity. In the bosonic case, the mutual information has the value $1$ in the vacuum and gradually approaches $2$ for massive black hole, while the negativity varies from zero to one. In the fermionic case, the mutual information shows similar behavior as the bosonic case, while the entanglement negativity has some finite value even for vacuum state and saturates to $1$ for massive black hole. Both mutual information and entanglement negativity for bosonic and fermionic field modes depend on the angular coordinates which is a consequence of the angle-dependent surface gravity at the soft hair horizon. The variation of surface gravity is justified by mapping its near horizon geometry to a shock wave Rindler background, where the soft hair function is closely related to the wave form factor.  It was observed that for vanishing soft hair function mutual information and entanglement negativity for both bososnic and fermionic field modes become equal to that of ordinary Schwarzschild black hole case. We have following comments: At first, the degradation of entanglement can be explained by the thermal decoherence seen by an accelerating observer, whose acceleration can be identified as the Unruh effect.  Nevertheless the Schwarzschild black hole is in fact thermally unstable due to its negative specific heat.  Author in the \cite{Wen:2021ahw} proposed the {\sl hair} temperature 
\begin{equation}
T_{\text {hair }}=\frac{\kappa}{2 \pi},
\end{equation}
which can be understood as the non-thermal temperature rather than usual Hawking temperature, driving subtle heat flow along angular direction by its gradient of soft hair function.  In this paper, we investigate an example of hair function $C \sim \cos^2{\theta}$.  Its gradient $\partial_\theta C \sim \sin{2\theta}$, which has a period $\pi/2$, consistent with the fluctuation of mutual information and negativity as shown in the Fig. \ref{mibfwh} to Fig. \ref{enffwh}.  Therefore according to our study, it is possible to indirectly {\sl observe} the soft hair by measurement of fluctuated entanglement at the Schwarzschild horizon. Secondly, the soft hair Schwarzschild metric in consideration is more
or less a classical background. If one believes in the entanglement nature of quantum gravity \cite{Krisnanda_2020}, we might instead consider an evaporating Schwarzschild black hole as an entangled state of the soft hair
metric and radiation, where the metric is supertranslated after each coded emission. We will
leave this for future study.

\section*{Acknowledgment}

This work is supported in part by the Taiwan’s Ministry of Science and Technology (109-2112-M-033-005-MY3) and the National Center for Theoretical Sciences (NCTS).

\begin{appendices}

\section*{Appendix} \label{appendix1}

Here we attempt to construct the map between hair function $C(z^A)$ and wave form factor $f(z^A)$ by studying the shock wave geometry.  We regard it as a product space of two-dimensional plane wave {\sl fiber} and a two-sphere {\sl base}.  Each patch $U_i$ on the base is an infinitesimal solid-angular region centered at angular coordinate $z^A_i$.  The plane wave fiber at each patch is coordinated by $(u_i,\hat{v_i})$, which are related to the Schwarzschild near-horizon coordinate $(t,x)$ by eq. \eqref{eqn:transform}.  Now consider in the region overlapped by two neighboring patches, say $U_1\cap U_2$, we may adopt the coordinate $u_1$ in the patch $U_1$ and $\hat{v_2}$ in the patch $U_2$ and write the shock wave fiber metric eq. \eqref{Min_sw} (neglect the sphere part) as
\begin{eqnarray}
ds^2 &= &-\mathrm{d} u_1 \mathrm{d} \hat{v}_2 - \partial_{z^A} f(z^A_i)\mathrm{d} u_1 \mathrm{d} z^A \nonumber\\ 
&= &e^{-(\kappa_1-\kappa_2)t}\bigg\{-x^2 \mathrm{d} t^2 + (\kappa_1\kappa_2)^{-1} \mathrm{d} x^2 - (\kappa_2^{-1}-\kappa_1^{-1}) x \mathrm{d} t \mathrm{d} x  \bigg\}- \partial_{z^A} f(z^A_i)\mathrm{d} u_1 \mathrm{d} z^A \nonumber\\
&\simeq &\bigg( 1 -(\kappa_1-\kappa_2)t + \cdots \bigg)\bigg\{-x^2 \mathrm{d} t^2 + (\kappa_1\kappa_2)^{-1} \mathrm{d} x^2 - (\kappa_2^{-1}-\kappa_1^{-1}) x \mathrm{d} t \mathrm{d} x  \bigg\}\nonumber\\
&-& \partial_{z^A} f(z^A_i)\mathrm{d} u_1 \mathrm{d} z^A   
\end{eqnarray}
For infinitesimally close patches, $|\kappa_1-\kappa_2| \ll 1$, this metric describes a near horizon geometry with surface gravity $(\kappa_1\kappa_2)^{1/2}$ if the last two cross terms were canceled.  To see it happen, we recall that by definition 
\begin{equation}
    \mathrm{d}\rho_s \simeq \mathrm{d}(\frac{M}{2}+Mx) = M\mathrm{d}x, \qquad \mathrm{d}\rho_s\big|_{\rho_s \simeq \frac{M}{2}}=\frac{\partial_{z^A} ||{\cal D}C||^2}{\rho_s}\mathrm{d}z^A = \frac{2}{M}\partial_A ||{\cal D}C||^2 \mathrm{d} z^A
\end{equation}
and from \eqref{eqn:transform}, 
\begin{eqnarray}
x = \sqrt{-\kappa_1\kappa_2 u_1\hat{v_2}} e^{(\kappa_1-\kappa_2)t/2} \simeq \sqrt{-\kappa_1\kappa_2 u_1\hat{v_2}} \big( 1 + (\kappa_1-\kappa_2)t/2 + \cdots \big) \\
\mathrm{d}t = \mathrm{d}\big(\frac{-1}{\kappa_1+\kappa_2} \ln{(\frac{-\kappa_1 u_1}{\kappa_2 \hat{v_2}})}\big) = -\frac{1}{\kappa_1+\kappa_2}\frac{\mathrm{d}u_1}{u_1} + \cdots
\end{eqnarray}
then the last two cross terms can be canceled if \footnote{One may instead use coordinate $\hat{v_1}$ in the patch $U_1$ and $u_2$ in the patch $U_2$.  The result is expected to be similar but with indices $1,2$ swapped.  Although there might be other construction of metric in the overlapped region, the result should be only different by order $O(\kappa_1-\kappa_2)$.}
\begin{equation}
    \frac{\kappa_1-\kappa_2}{\kappa_1+\kappa_2}\sqrt{\frac{-\hat{v_2}^*}{u_1^*\kappa_1\kappa_2}}\frac{2}{M^2}\partial_{z^A}||{\cal D}C||^2 = \partial_{z^A}f.
\end{equation}
Note we have assigned $u_1=u_1^*, \hat{v_2}=\hat{v_2}^*$ at which the shock wave forms the black hole.  In this way we have explicitly constructed the {\sl differential} relation between soft hair function and wave form factor.

\end{appendices}


\bibliographystyle{JHEP}

\bibliography{entsw}

\providecommand{\href}[2]{#2}\begingroup\raggedright\begin{thebibliography}{10}

\bibitem{Calabrese:2004eu}
P.~Calabrese and J.~L. Cardy, \emph{{Entanglement entropy and quantum field
  theory}}, \href{https://doi.org/10.1088/1742-5468/2004/06/P06002}{\emph{J.
  Stat. Mech.} {\bfseries 0406} (2004) P06002},
  [\href{https://arxiv.org/abs/hep-th/0405152}{{\ttfamily hep-th/0405152}}].

\bibitem{Calabrese:2005zw}
P.~Calabrese and J.~L. Cardy, \emph{{Entanglement entropy and quantum field
  theory: A Non-technical introduction}},
  \href{https://doi.org/10.1142/S021974990600192X}{\emph{Int. J. Quant. Inf.}
  {\bfseries 4} (2006) 429},
  [\href{https://arxiv.org/abs/quant-ph/0505193}{{\ttfamily
  quant-ph/0505193}}].

\bibitem{Calabrese:2009qy}
P.~Calabrese and J.~Cardy, \emph{{Entanglement entropy and conformal field
  theory}}, \href{https://doi.org/10.1088/1751-8113/42/50/504005}{\emph{J.
  Phys.} {\bfseries A42} (2009) 504005},
  [\href{https://arxiv.org/abs/0905.4013}{{\ttfamily 0905.4013}}].

\bibitem{Ryu:2006bv}
S.~Ryu and T.~Takayanagi, \emph{{Holographic derivation of entanglement entropy
  from AdS/CFT}},
  \href{https://doi.org/10.1103/PhysRevLett.96.181602}{\emph{Phys. Rev. Lett.}
  {\bfseries 96} (2006) 181602},
  [\href{https://arxiv.org/abs/hep-th/0603001}{{\ttfamily hep-th/0603001}}].

\bibitem{Ryu:2006ef}
S.~Ryu and T.~Takayanagi, \emph{{Aspects of Holographic Entanglement Entropy}},
  \href{https://doi.org/10.1088/1126-6708/2006/08/045}{\emph{JHEP} {\bfseries
  08} (2006) 045}, [\href{https://arxiv.org/abs/hep-th/0605073}{{\ttfamily
  hep-th/0605073}}].

\bibitem{Takayanagi:2012kg}
T.~Takayanagi, \emph{{Entanglement Entropy from a Holographic Viewpoint}},
  \href{https://doi.org/10.1088/0264-9381/29/15/153001}{\emph{Class. Quant.
  Grav.} {\bfseries 29} (2012) 153001},
  [\href{https://arxiv.org/abs/1204.2450}{{\ttfamily 1204.2450}}].

\bibitem{Nishioka:2009un}
T.~Nishioka, S.~Ryu and T.~Takayanagi, \emph{{Holographic Entanglement Entropy:
  An Overview}}, \href{https://doi.org/10.1088/1751-8113/42/50/504008}{\emph{J.
  Phys.} {\bfseries A42} (2009) 504008},
  [\href{https://arxiv.org/abs/0905.0932}{{\ttfamily 0905.0932}}].

\bibitem{Nishioka:2018khk}
T.~Nishioka, \emph{Entanglement entropy: Holography and renormalization group},
  \href{https://doi.org/10.1103/revmodphys.90.035007}{\emph{Reviews of Modern
  Physics} {\bfseries 90} (sep, 2018) }.

\bibitem{e12112244}
M.~Cadoni and M.~Melis, \emph{Entanglement entropy of ads black holes},
  \href{https://doi.org/10.3390/e12112244}{\emph{Entropy} {\bfseries 12} (2010)
  2244--2267}.

\bibitem{Blanco:2013joa}
D.~D. Blanco, H.~Casini, L.-Y. Hung and R.~C. Myers, \emph{{Relative Entropy
  and Holography}}, \href{https://doi.org/10.1007/JHEP08(2013)060}{\emph{JHEP}
  {\bfseries 08} (2013) 060},
  [\href{https://arxiv.org/abs/1305.3182}{{\ttfamily 1305.3182}}].

\bibitem{Fischler2013}
W.~Fischler and S.~Kundu, \emph{Strongly coupled gauge theories: high and low
  temperature behavior of non-local observables},
  \href{https://doi.org/10.1007/JHEP05(2013)098}{\emph{Journal of High Energy
  Physics} {\bfseries 2013} (May, 2013) 98}.

\bibitem{Fischler:2012uv}
W.~Fischler, A.~Kundu and S.~Kundu, \emph{{Holographic Mutual Information at
  Finite Temperature}},
  \href{https://doi.org/10.1103/PhysRevD.87.126012}{\emph{Phys. Rev.}
  {\bfseries D87} (2013) 126012},
  [\href{https://arxiv.org/abs/1212.4764}{{\ttfamily 1212.4764}}].

\bibitem{Chaturvedi:2016kbk}
P.~Chaturvedi, V.~Malvimat and G.~Sengupta, \emph{{Entanglement thermodynamics
  for charged black holes}},
  \href{https://doi.org/10.1103/PhysRevD.94.066004}{\emph{Phys. Rev.}
  {\bfseries D94} (2016) 066004},
  [\href{https://arxiv.org/abs/1601.00303}{{\ttfamily 1601.00303}}].

\bibitem{Fursaev:2006ih}
D.~V. Fursaev, \emph{{Proof of the holographic formula for entanglement
  entropy}}, \href{https://doi.org/10.1088/1126-6708/2006/09/018}{\emph{JHEP}
  {\bfseries 09} (2006) 018},
  [\href{https://arxiv.org/abs/hep-th/0606184}{{\ttfamily hep-th/0606184}}].

\bibitem{Headrick:2010zt}
M.~Headrick, \emph{{Entanglement Renyi entropies in holographic theories}},
  \href{https://doi.org/10.1103/PhysRevD.82.126010}{\emph{Phys. Rev.}
  {\bfseries D82} (2010) 126010},
  [\href{https://arxiv.org/abs/1006.0047}{{\ttfamily 1006.0047}}].

\bibitem{Faulkner:2013yia}
T.~Faulkner, \emph{{The Entanglement Renyi Entropies of Disjoint Intervals in
  AdS/CFT}},  \href{https://arxiv.org/abs/1303.7221}{{\ttfamily 1303.7221}}.

\bibitem{Casini:2011kv}
H.~Casini, M.~Huerta and R.~C. Myers, \emph{{Towards a derivation of
  holographic entanglement entropy}},
  \href{https://doi.org/10.1007/JHEP05(2011)036}{\emph{JHEP} {\bfseries 05}
  (2011) 036}, [\href{https://arxiv.org/abs/1102.0440}{{\ttfamily 1102.0440}}].

\bibitem{Lewkowycz:2013nqa}
A.~Lewkowycz and J.~Maldacena, \emph{{Generalized gravitational entropy}},
  \href{https://doi.org/10.1007/JHEP08(2013)090}{\emph{JHEP} {\bfseries 08}
  (2013) 090}, [\href{https://arxiv.org/abs/1304.4926}{{\ttfamily 1304.4926}}].

\bibitem{Hubeny:2007xt}
V.~E. Hubeny, M.~Rangamani and T.~Takayanagi, \emph{{A Covariant holographic
  entanglement entropy proposal}},
  \href{https://doi.org/10.1088/1126-6708/2007/07/062}{\emph{JHEP} {\bfseries
  07} (2007) 062}, [\href{https://arxiv.org/abs/0705.0016}{{\ttfamily
  0705.0016}}].

\bibitem{Dong:2016hjy}
X.~Dong, A.~Lewkowycz and M.~Rangamani, \emph{{Deriving covariant holographic
  entanglement}}, \href{https://doi.org/10.1007/JHEP11(2016)028}{\emph{JHEP}
  {\bfseries 11} (2016) 028},
  [\href{https://arxiv.org/abs/1607.07506}{{\ttfamily 1607.07506}}].

\bibitem{PhysRevA.65.032314}
G.~Vidal and R.~F. Werner, \emph{Computable measure of entanglement},
  \href{https://doi.org/10.1103/PhysRevA.65.032314}{\emph{Phys. Rev. A}
  {\bfseries 65} (Feb, 2002) 032314}.

\bibitem{Plenio:2005cwa}
M.~B. Plenio, \emph{{Logarithmic Negativity: A Full Entanglement Monotone That
  is not Convex}},
  \href{https://doi.org/10.1103/PhysRevLett.95.090503}{\emph{Phys. Rev. Lett.}
  {\bfseries 95} (2005) 090503},
  [\href{https://arxiv.org/abs/quant-ph/0505071}{{\ttfamily
  quant-ph/0505071}}].

\bibitem{Calabrese:2012ew}
P.~Calabrese, J.~Cardy and E.~Tonni, \emph{{Entanglement negativity in quantum
  field theory}},
  \href{https://doi.org/10.1103/PhysRevLett.109.130502}{\emph{Phys. Rev. Lett.}
  {\bfseries 109} (2012) 130502},
  [\href{https://arxiv.org/abs/1206.3092}{{\ttfamily 1206.3092}}].

\bibitem{Calabrese:2012nk}
P.~Calabrese, J.~Cardy and E.~Tonni, \emph{{Entanglement negativity in extended
  systems: A field theoretical approach}},
  \href{https://doi.org/10.1088/1742-5468/2013/02/P02008}{\emph{J. Stat. Mech.}
  {\bfseries 1302} (2013) P02008},
  [\href{https://arxiv.org/abs/1210.5359}{{\ttfamily 1210.5359}}].

\bibitem{Calabrese:2014yza}
P.~Calabrese, J.~Cardy and E.~Tonni, \emph{{Finite temperature entanglement
  negativity in conformal field theory}},
  \href{https://doi.org/10.1088/1751-8113/48/1/015006}{\emph{J. Phys.}
  {\bfseries A48} (2015) 015006},
  [\href{https://arxiv.org/abs/1408.3043}{{\ttfamily 1408.3043}}].

\bibitem{Rangamani:2014ywa}
M.~Rangamani and M.~Rota, \emph{{Comments on Entanglement Negativity in
  Holographic Field Theories}},
  \href{https://doi.org/10.1007/JHEP10(2014)060}{\emph{JHEP} {\bfseries 10}
  (2014) 060}, [\href{https://arxiv.org/abs/1406.6989}{{\ttfamily 1406.6989}}].

\bibitem{Perlmutter:2015vma}
E.~Perlmutter, M.~Rangamani and M.~Rota, \emph{{Central Charges and the Sign of
  Entanglement in 4D Conformal Field Theories}},
  \href{https://doi.org/10.1103/PhysRevLett.115.171601}{\emph{Phys. Rev. Lett.}
  {\bfseries 115} (2015) 171601},
  [\href{https://arxiv.org/abs/1506.01679}{{\ttfamily 1506.01679}}].

\bibitem{Chaturvedi:2016rcn}
P.~Chaturvedi, V.~Malvimat and G.~Sengupta, \emph{{Holographic Quantum
  Entanglement Negativity}},
  \href{https://doi.org/10.1007/JHEP05(2018)172}{\emph{JHEP} {\bfseries 05}
  (2018) 172}, [\href{https://arxiv.org/abs/1609.06609}{{\ttfamily
  1609.06609}}].

\bibitem{Chaturvedi:2016opa}
P.~Chaturvedi, V.~Malvimat and G.~Sengupta, \emph{{Covariant holographic
  entanglement negativity}},
  \href{https://doi.org/10.1140/epjc/s10052-018-6259-1}{\emph{Eur. Phys. J.}
  {\bfseries C78} (2018) 776},
  [\href{https://arxiv.org/abs/1611.00593}{{\ttfamily 1611.00593}}].

\bibitem{Malvimat:2017yaj}
G.~Sengupta and V.~Malvimat, \emph{Entanglement negativity at large central
  charge}, \href{https://doi.org/10.1103/physrevd.103.106003}{\emph{Physical
  Review D} {\bfseries 103} (May, 2021) }.

\bibitem{Chaturvedi:2016rft}
P.~Chaturvedi, V.~Malvimat and G.~Sengupta, \emph{{Entanglement negativity,
  Holography and Black holes}},
  \href{https://doi.org/10.1140/epjc/s10052-018-5969-8}{\emph{Eur. Phys. J.}
  {\bfseries C78} (2018) 499},
  [\href{https://arxiv.org/abs/1602.01147}{{\ttfamily 1602.01147}}].

\bibitem{Jain_2019hen}
P.~Jain, V.~Malvimat, S.~Mondal and G.~Sengupta, \emph{Holographic entanglement
  negativity conjecture for adjacent intervals in {AdS}3/{CFT}2},
  \href{https://doi.org/10.1016/j.physletb.2019.04.037}{\emph{Physics Letters
  B} {\bfseries 793} (Jun, 2019) 104--109}.

\bibitem{Jain:2017uhe}
P.~Jain, V.~Malvimat, S.~Mondal and G.~Sengupta, \emph{Covariant holographic
  entanglement negativity for adjacent subsystems in {AdS}3/{CFT}2},
  \href{https://doi.org/10.1016/j.nuclphysb.2019.114683}{\emph{Nuclear Physics
  B} {\bfseries 945} (Aug, 2019) 114683}.

\bibitem{Jain:2017xsu}
P.~Jain, V.~Malvimat, S.~Mondal and G.~Sengupta, \emph{{Holographic
  entanglement negativity for adjacent subsystems in AdS$_{d+1}$/CFT$_{d}$}},
  \href{https://doi.org/10.1140/epjp/i2018-12113-0}{\emph{Eur. Phys. J. Plus}
  {\bfseries 133} (2018) 300},
  [\href{https://arxiv.org/abs/1708.00612}{{\ttfamily 1708.00612}}].

\bibitem{Jain:2018bai}
P.~Jain, V.~Malvimat, S.~Mondal and G.~Sengupta, \emph{{Holographic
  Entanglement Negativity for Conformal Field Theories with a Conserved
  Charge}}, \href{https://doi.org/10.1140/epjc/s10052-018-6383-y}{\emph{Eur.
  Phys. J.} {\bfseries C78} (2018) 908},
  [\href{https://arxiv.org/abs/1804.09078}{{\ttfamily 1804.09078}}].

\bibitem{Malvimat:2018txq}
V.~Malvimat, S.~Mondal, B.~Paul and G.~Sengupta, \emph{Holographic entanglement
  negativity for disjoint intervals in ads3/cft2},
  \href{https://doi.org/10.1140/epjc/s10052-019-6693-8}{\emph{The European
  Physical Journal C} {\bfseries 79} (Mar, 2019) }.

\bibitem{Malvimat:2018ood}
V.~Malvimat, S.~Mondal, B.~Paul and G.~Sengupta, \emph{Covariant holographic
  entanglement negativity for disjoint intervals in ads3/cft2},
  \href{https://doi.org/10.1140/epjc/s10052-019-7032-9}{\emph{The European
  Physical Journal C} {\bfseries 79} (Jun, 2019) }.

\bibitem{Malvimat_2019}
V.~Malvimat, S.~Mondal and G.~Sengupta, \emph{Time evolution of entanglement
  negativity from black hole interiors},
  \href{https://doi.org/10.1007/jhep05(2019)183}{\emph{Journal of High Energy
  Physics} {\bfseries 2019} (May, 2019) }.

\bibitem{basak2020minimal}
J.~K. Basak, V.~Malvimat, H.~Parihar, B.~Paul and G.~Sengupta, \emph{On minimal
  entanglement wedge cross section for holographic entanglement negativity},
  2020.
\newblock 10.48550/ARXIV.2002.10272.

\bibitem{basak2021islands}
J.~K. Basak, D.~Basu, V.~Malvimat, H.~Parihar and G.~Sengupta, \emph{Islands
  for entanglement negativity},
  \href{https://doi.org/10.21468/scipostphys.12.1.003}{\emph{{SciPost} Physics}
  {\bfseries 12} (Jan, 2022) }.

\bibitem{basak2021page}
J.~K. Basak, D.~Basu, V.~Malvimat, H.~Parihar and G.~Sengupta, \emph{Page curve
  for entanglement negativity through geometric evaporation},
  \href{https://doi.org/10.21468/scipostphys.12.1.004}{\emph{{SciPost} Physics}
  {\bfseries 12} (Jan, 2022) }.

\bibitem{afrasiar2021holographic}
M.~Afrasiar, J.~K. Basak, V.~Raj and G.~Sengupta, \emph{Holographic
  entanglement negativity for disjoint subsystems in conformal field theories
  with a conserved charge},  2021.

\bibitem{basu2021entanglement}
D.~Basu, A.~Chandra, V.~Raj and G.~Sengupta, \emph{Entanglement wedge in flat
  holography and entanglement negativity},  2021.

\bibitem{mondal2021holographic}
S.~Mondal, B.~Paul, G.~Sengupta and P.~Sharma, \emph{Holographic entanglement
  negativity for a single subsystem in conformal field theories with a
  conserved charge},  2021.

\bibitem{Dong_2021}
X.~Dong, X.-L. Qi and M.~Walter, \emph{Holographic entanglement negativity and
  replica symmetry breaking},
  \href{https://doi.org/10.1007/jhep06(2021)024}{\emph{Journal of High Energy
  Physics} {\bfseries 2021} (Jun, 2021) }.

\bibitem{Kudler_Flam_2020}
J.~Kudler-Flam, H.~Shapourian and S.~Ryu, \emph{The negativity contour: a
  quasi-local measure of entanglement for mixed states},
  \href{https://doi.org/10.21468/scipostphys.8.4.063}{\emph{SciPost Physics}
  {\bfseries 8} (Apr, 2020) }.

\bibitem{Kudler_Flam_2019}
J.~Kudler-Flam and S.~Ryu, \emph{Entanglement negativity and minimal
  entanglement wedge cross sections in holographic theories},
  \href{https://doi.org/10.1103/physrevd.99.106014}{\emph{Physical Review D}
  {\bfseries 99} (May, 2019) }.

\bibitem{PhysRevLett.91.180404}
P.~M. Alsing and G.~J. Milburn, \emph{Teleportation with a uniformly
  accelerated partner},
  \href{https://doi.org/10.1103/PhysRevLett.91.180404}{\emph{Phys. Rev. Lett.}
  {\bfseries 91} (Oct, 2003) 180404}.

\bibitem{Alsing_2004}
P.~M. Alsing, D.~McMahon and G.~J. Milburn, \emph{Teleportation in a
  non-inertial frame},
  \href{https://doi.org/10.1088/1464-4266/6/8/033}{\emph{Journal of Optics B:
  Quantum and Semiclassical Optics} {\bfseries 6} (jul, 2004) S834--S843}.

\bibitem{PhysRevLett.95.120404}
I.~Fuentes-Schuller and R.~B. Mann, \emph{Alice falls into a black hole:
  Entanglement in noninertial frames},
  \href{https://doi.org/10.1103/PhysRevLett.95.120404}{\emph{Phys. Rev. Lett.}
  {\bfseries 95} (Sep, 2005) 120404}.

\bibitem{PhysRevA.74.032326}
P.~M. Alsing, I.~Fuentes-Schuller, R.~B. Mann and T.~E. Tessier,
  \emph{Entanglement of dirac fields in noninertial frames},
  \href{https://doi.org/10.1103/PhysRevA.74.032326}{\emph{Phys. Rev. A}
  {\bfseries 74} (Sep, 2006) 032326}.

\bibitem{Hwang_2011}
M.-R. Hwang, D.~Park and E.~Jung, \emph{Tripartite entanglement in a
  noninertial frame},
  \href{https://doi.org/10.1103/physreva.83.012111}{\emph{Physical Review A}
  {\bfseries 83} (Jan, 2011) }.

\bibitem{shamirzai2011tripartite}
M.~Shamirzai, B.~N. Esfahani and M.~Soltani, \emph{Tripartite entanglements in
  non-inertial frames},  2011.

\bibitem{Alsing_2003}
P.~M. Alsing and G.~J. Milburn, \emph{Teleportation with a uniformly
  accelerated partner},
  \href{https://doi.org/10.1103/physrevlett.91.180404}{\emph{Physical Review
  Letters} {\bfseries 91} (Oct, 2003) }.

\bibitem{PhysRevA.81.032320}
E.~Mart\'{\i}n-Mart\'{\i}nez and J.~Le\'on, \emph{Quantum correlations through
  event horizons: Fermionic versus bosonic entanglement},
  \href{https://doi.org/10.1103/PhysRevA.81.032320}{\emph{Phys. Rev. A}
  {\bfseries 81} (Mar, 2010) 032320}.

\bibitem{PhysRevA.80.042318}
E.~Mart\'{\i}n-Mart\'{\i}nez and J.~Le\'on, \emph{Fermionic entanglement that
  survives a black hole},
  \href{https://doi.org/10.1103/PhysRevA.80.042318}{\emph{Phys. Rev. A}
  {\bfseries 80} (Oct, 2009) 042318}.

\bibitem{PhysRevA.82.042332}
D.~E. Bruschi, J.~Louko, E.~Mart\'{\i}n-Mart\'{\i}nez, A.~Dragan and
  I.~Fuentes, \emph{Unruh effect in quantum information beyond the single-mode
  approximation}, \href{https://doi.org/10.1103/PhysRevA.82.042332}{\emph{Phys.
  Rev. A} {\bfseries 82} (Oct, 2010) 042332}.

\bibitem{Hwang_2012}
M.-R. Hwang, E.~Jung and D.~Park, \emph{Three-tangle in non-inertial frame},
  \href{https://doi.org/10.1088/0264-9381/29/22/224004}{\emph{Classical and
  Quantum Gravity} {\bfseries 29} (oct, 2012) 224004}.

\bibitem{PhysRevD.14.870}
W.~G. Unruh, \emph{Notes on black-hole evaporation},
  \href{https://doi.org/10.1103/PhysRevD.14.870}{\emph{Phys. Rev. D} {\bfseries
  14} (Aug, 1976) 870--892}.

\bibitem{Hawking:2016msc}
S.~W. Hawking, M.~J. Perry and A.~Strominger, \emph{{Soft Hair on Black
  Holes}}, \href{https://doi.org/10.1103/PhysRevLett.116.231301}{\emph{Phys.
  Rev. Lett.} {\bfseries 116} (2016) 231301},
  [\href{https://arxiv.org/abs/1601.00921}{{\ttfamily 1601.00921}}].

\bibitem{Strominger:2017aeh}
A.~Strominger, \emph{Black hole information revisited},  2017.
\newblock 10.48550/ARXIV.1706.07143.

\bibitem{Perry:2020tts}
M.~J. Perry, \emph{{Black hole entropy from soft hair}},
  \href{https://doi.org/10.1142/S0218271820300128}{\emph{Int. J. Mod. Phys. D}
  {\bfseries 29} (2020) 2030012}.

\bibitem{Wen:2021ahw}
W.-Y. Wen, \emph{{Dressed tunneling in soft hair}},
  \href{https://doi.org/10.1016/j.physletb.2021.136578}{\emph{Phys. Lett. B}
  {\bfseries 820} (2021) 136578},
  [\href{https://arxiv.org/abs/2103.00516}{{\ttfamily 2103.00516}}].

\bibitem{DRAY1985173}
T.~Dray and G.~{'t Hooft}, \emph{The gravitational shock wave of a massless
  particle},
  \href{https://doi.org/https://doi.org/10.1016/0550-3213(85)90525-5}{\emph{Nuclear
  Physics B} {\bfseries 253} (1985) 173--188}.

\bibitem{Comp_re_2019}
G.~Compère, J.~Long and M.~Riegler, \emph{Invariance of unruh and hawking
  radiation under matter-induced supertranslations},
  \href{https://doi.org/10.1007/jhep05(2019)053}{\emph{Journal of High Energy
  Physics} {\bfseries 2019} (May, 2019) }.

\bibitem{Camanho_2016}
X.~O. Camanho, J.~D. Edelstein, J.~Maldacena and A.~Zhiboedov, \emph{Causality
  constraints on corrections to the graviton three-point coupling},
  \href{https://doi.org/10.1007/jhep02(2016)020}{\emph{Journal of High Energy
  Physics} {\bfseries 2016} (Feb, 2016) }.

\bibitem{Compere:2016hzt}
G.~Comp\`ere and J.~Long, \emph{{Classical static final state of collapse with
  supertranslation memory}},
  \href{https://doi.org/10.1088/0264-9381/33/19/195001}{\emph{Class. Quant.
  Grav.} {\bfseries 33} (2016) 195001},
  [\href{https://arxiv.org/abs/1602.05197}{{\ttfamily 1602.05197}}].

\bibitem{Comp_re_2016}
G.~Comp{\`{e} }re and J.~Long, \emph{Classical static final state of collapse
  with supertranslation memory},
  \href{https://doi.org/10.1088/0264-9381/33/19/195001}{\emph{Classical and
  Quantum Gravity} {\bfseries 33} (sep, 2016) 195001}.

\bibitem{Lin_2020}
F.-L. Lin and S.~Takeuchi, \emph{Hawking flux from a black hole with nonlinear
  supertranslation hair},
  \href{https://doi.org/10.1103/physrevd.102.044004}{\emph{Physical Review D}
  {\bfseries 102} (aug, 2020) }.

\bibitem{2008HR}
Q.~Pan and J.~Jing, \emph{Hawking radiation, entanglement, and teleportation in
  the background of an asymptotically flat static black hole},
  \href{https://doi.org/10.1103/physrevd.78.065015}{\emph{Physical Review D}
  {\bfseries 78} (Sep, 2008) }.

\bibitem{Krisnanda_2020}
T.~Krisnanda, G.~Y. Tham, M.~Paternostro and T.~Paterek, \emph{Observable
  quantum entanglement due to gravity},
  \href{https://doi.org/10.1038/s41534-020-0243-y}{\emph{NPJ Quantum
  Information} {\bfseries 6} (Jan, 2020) }.

\end{thebibliography}\endgroup

\end{document}